\def\sqr#1#2{{\vcenter{\vbox{\hrule height.#2pt\hbox{\vrule width.#2pt 
height#1pt \kern#1pt \vrule width.#2pt}\hrule height.#2pt}}}}
\def\d{\partial}

\def\=d{\,{\buildrel\rm def\over =}\,}

\documentclass[12pt]{article}
\usepackage{amssymb}\usepackage{amsmath}\usepackage{amsthm}

\newcommand{\beq}{\begin{equation}}
\newcommand{\eeq}{\end{equation}}
\newcommand{\bg}{\begin{gather}}
\newcommand{\eg}{\end{gather}}
\newcommand{\CC}{\mathbb C}
\newcommand{\RR}{\mathbb R}
\newcommand{\ZZ}{\mathbb Z}
\newcommand{\NN}{\mathbb N}

\newcommand{\supp}{{\rm supp\>}}

\newtheorem{prop}{Proposition}

\newtheorem{cor}[prop]{Corollary}
\newtheorem{lemma}[prop]{Lemma}

\begin{document}
\title{Proof of perturbative gauge invariance for tree 
diagrams to all orders}
\author{Michael D\"utsch \\[2mm] 
Institut f\"ur Theoretische Physik\\
Universit\"at Z\"urich\\
CH-8057 Z\"urich, Switzerland\\
{\tt \small duetsch@physik.unizh.ch}\\[2mm]}

\date{}
\maketitle
\begin{abstract}
It is proved that classical BRS-invariance of the Lagrangian implies perturbative gauge invariance 
for tree diagrams to all orders. The proof applies in particular to 
the Einstein Hilbert Largrangian of gravity.

\end{abstract}

\tableofcontents
\section{Introduction}\setcounter{equation}{0}
One of the greatest challenges of present-day quantum field theory (QFT) is the search 
for a quantum theory of gravity. In this context revolutionary approaches are intensively studied, 
e.g.~non-commutative space-times, string theory and loop quantum gravity. Since this paper
is related to BRS-symmetry \cite{BRS}, we only mention that a BRS-formulation of gravity
was given in \cite{DRM,Dix,TN} and that the structure of the possible anomalies has been 
worked out by cohomological methods, see e.g.~ \cite{BDK,BBH}.

The main result of this paper is much more modest: we prove
{\it perturbative gauge invariance} (PGI) \cite{DHKS,Scharf,S-wiley,G,stora}
(which is a condition in perturbative QFT that is related to BRS-invariance) for gravity,
but our result is restricted to {\it tree diagrams}. Since PGI for tree diagrams
(PGI-tree) is equivalent to PGI in {\it classical} field theory (cf.~Appendix B of
\cite{DF2} and Sect.~2), our result is actually a statement for classical gravity.
However, it is also a justification to use PGI for the construction of a perturbative QFT
for spin-2 gauge fields - a project started in \cite{Schorn,SW,Grillo,S-wiley,G:grav,GS}.

In the latter the requirement of PGI-tree has been used to determine the possible
interactions of massless spin-2
gauge fields \cite{Schorn,SW,S-wiley}. Making a polynomial 
ansatz for the interaction ${\cal L}=\sum_{n=1}^\infty\kappa^n\,{\cal L}^{(n)}$
(where $\kappa$ is the coupling constant), it has been worked out that the most 
general solutions for ${\cal L}^{(1)}$ and ${\cal L}^{(2)}$ agree with the 
corresponding terms of the Einstein-Hilbert Lagrangian of gravity 
up to physically irrelevant terms, see \cite{SW,S-wiley}.
But continuing this procedure to higher orders the amount of computational
work increases strongly and, due to the non-renormalizability of spin-2
gauge fields, one never comes to an end. That is, violations of PGI-tree
can appear to arbitrary high orders and, if PGI-tree can be fulfilled, it is not clear, 
that the general solution for ${\cal L}^{(k)},\>k\geq 3$, agrees with the corresponding 
term of the Einstein-Hilbert Lagrangian. The {\it main purpose of this paper is to prove 
that the Einstein-Hilbert Lagrangian, completed by a gauge fixing and a 
Faddeev-Popov ghost term} (we follow \cite{KO}), {\it yields a solution of
PGI-tree to all orders}.

To a large extent we formulate the proof independently of the model. However,
the applications to massless and massive spin-1 fields (Sects.~4.1 and 4.2)
are only of pedagogical value: for {\it renormalizable models} a violation of PGI-tree
can usually\footnote{The statement is valid for renormalizable models in $4$ 
dimensions with the property that all terms of the interaction ${\cal L}$
are at least of third order in the basic fields.}
appear only up to third order (as can easily be seen by power counting), and
using this fact, PGI-tree has been proved by explicit computation of the lowest orders
for various spin-1 models \cite{DHKS,DS,G}.

By {\it perturbative gauge invariance} we mean the following condition. Let a free 
quantum gauge theory (i.e.~a free Lagrangian ${\cal L}^{(0)}$) and the corresponding 
free BRS-transformation $s_0$ be given and let ${\cal L}^{(0)}$ be 
$s_0$ invariant, that is
$s_0\,{\cal L}^{(0)}=-\d_\mu I^{(0)\,\mu}$ for some field polynomial $I^{(0)\,\mu}$. 
In addition let $j^{(0)\,\mu}$ be the corresponding conserved Noether current,
$\d_\mu j^{(0)\,\mu}_{S_0}=0$, and let $Q$ be the corresponding charge: 
$Q=\int d^3x\,j_{S_0}^{(0)\,0}(x^0,\vec{x})$ (``free BRS-charge''). 
(The lower index $S_0$ signifies always
that we mean the 'on-shell fields', i.e.~the free field equations are valid; for 
a precise formulation see \cite{DF2,DF3} or Appendix A.) 
PGI requires that to an interaction ${\cal L}^{(1)}$ there exists a 
Lorentz vector ${\cal L}_1^{(1)\,\nu}$ and a normalization of the time ordered products 
$T^N$ such that
\beq
[Q,T_{S_0}^N\bigl({\cal L}^{(1)}(x_1)...{\cal L}^{(1)}(x_n)\bigr)]=i\sum_{l=1}^n\d_\nu^{x_l}\>
T_{S_0}^N\bigl({\cal L}^{(1)}(x_1)...{\cal L}_1^{(1)\,\nu}(x_l)...{\cal L}^{(1)}(x_n)\bigr)\label{PGI}
\eeq
(where $[\,\cdot\,,\,\cdot\,]$ denotes the commutator with respect to the $\star$ product).
The upper index $(1)$ of  ${\cal L}^{(1)}$ signifies that we mean the term of first order 
(in the coupling constant $\kappa$) of the total interaction ${\cal L}=\sum_{n=1}^\infty
\kappa^n\,{\cal L}^{(n)}$ and similarly ${\cal L}_1^{(1)\,\nu}$ is the term of first
order of the total ``$Q$-vertex'' ${\cal L}_1^{\nu}=\sum_{n=1}^\infty
\kappa^n\,{\cal L}_1^{(n)\,\nu}$. Higher order terms ${\cal L}^{(n)}$ (${\cal L}_1^{(n)\,\nu}$
resp.), $n\geq 2$, are absorbed in a finite renormalization $T({\cal L}^{(1)}...{\cal L}^{(1)})
\rightarrow T^N({\cal L}^{(1)}...{\cal L}^{(1)})$ ($T({\cal L}^{(1)}...{\cal L}_1^{(1)\,\nu}...)
\rightarrow T^N({\cal L}^{(1)}...{\cal L}_1^{(1)\,\nu}...)$ resp.) of tree diagrams. This is 
always possible, as shown in Sect.~2.

PGI plays two different roles:
\begin{itemize}
\item [(i)] PGI-tree restricts the interaction ${\cal L}=\sum_{n=1}^\infty\kappa^n\,{\cal L}^{(n)}$
strongly, as mentioned above.
\item  [(ii)] For loop diagrams it is a highly non-trivial (re)normalization condition,
which cannot always be fulfilled, e.g.~in case of the axial anomaly.
\end{itemize}

The motivations to require PGI are the following.
\begin{itemize}
\item[(A)] In purely massive theories the adiabatic limit exists \cite{EG1}, i.e.~there is an $S$-matrix
\begin{equation}
S={\bf 1}+\lim_{g\to 1}\sum_{n=1}^\infty\frac{i^n}{n!}\int dx_1...dx_n\,g(x_1)...g(x_n)\,
T_{S_0}^N\bigl({\cal L}^{(1)}(x_1)...{\cal L}^{(1)}(x_n)\bigr)\ .\label{S}
\end{equation}
Kugo and Ojima \cite{KO1,KO} have shown that in the adiabatic limit the physical
Hilbert space ${\cal H}$ can be expressed in terms of the free BRS-charge $Q$:
\begin{equation}
{\cal H}=\frac{{\rm ker}\>Q}{{\rm ran}\>Q}\ .
\end{equation}
PGI implies $[Q,S]=0$ and, hence, $S$ is well defined on ${\cal H}$. That is,
PGI is a sufficient condition for the quantization of purely massive gauge theories and,
as shown in \cite{DSchroer}, it is even almost necessary for this purpose.
\item[(B)] Since PGI is well defined also for theories in which the adiabatic limit 
does not exist, PGI-tree can be used to derive the
interaction ${\cal L}$ for all kinds of gauge theories. Making a polynomial 
ansatz for ${\cal L}$, PGI-tree and
some obvious requirements (e.g.~Lorentz invariance, ghost number zero and in
case of spin-1 fields renormalizability) determine ${\cal L}$ to a far extent. We
recall the highlights (besides the already mentioned derivation of the Einstein-Hilbert
Lagrangian). 
\begin{itemize}
\item The Lie algebraic structure of spin-1 fields needs not to be put in, 
it can be derived in this way \cite{stora}.
\item For non-Abelian massive spin-1 theories it is impossible to satisfy these
requirements for a model with only gauge fields and ghosts (fermionic and 
bosonic). The inclusion of additional {\it physical} scalar fields 
(corresponding to Higgs fields) yields a solution \cite{DS}. 
\end{itemize}
\end{itemize}

In this paper we proceed in the direction opposite to (B): we assume that a Lagrangian
${\cal L}_{\rm total}=\sum_{n=0}^\infty\kappa^n\,{\cal L}^{(n)}$ and a 
BRS-transformation $s=\sum_{n=0}^\infty\kappa^n\,s_n$ (of the interacting fields) 
\cite{BRS} are given and that ${\cal L}_{\rm total}$ is BRS-invariant: 
$s\,{\cal L}_{\rm total}=-\d_\mu I^{\mu}$ for some 
formal power series $I^{\mu}$. We prove that this assumption implies PGI-tree.

Our proof of PGI-tree relies strongly on results which have been found in
\cite{DF}, \cite{BD} and \cite{DF2}. Namely, let us start with the local conservation of the
BRS-current,
\begin{gather}
\d_\mu^x\,T_{S_0}^N\bigl(j^{(0)\,\mu}(x),{\cal L}^{(1)}(x_1)...{\cal L}^{(1)}(x_n)\bigr)=\notag\\
-i\sum_{l=1}^n
\d_\nu^{x_l}\,\Bigl(\delta(x-x_l)\>T_{S_0}^N\bigl({\cal L}^{(1)}(x_1)...{\cal L}_1^{(1)\,\nu}
(x_l)...{\cal L}^{(1)}(x_n)\bigr)\Bigr)\label{localcurrconserv}
\end{gather}
where the higher order terms $j^{(n)}\,,\>n\geq 1$, of the total BRS-current
$j^{\mu}=\sum_{n=0}^\infty\kappa^n\,j^{(n)\,\mu}$ are absorbed in a renormalization
$T(j^{(0)},{\cal L}^{(1)}...)\rightarrow T^N(j^{(0)},{\cal L}^{(1)}...)$ of tree diagrams
(see Sect.~2). In Appendix B of \cite{DF} and in \cite{BD} it is proved that by 
smearing out (\ref{localcurrconserv}) with a test function $f(x)$
which satisfies $f\vert_{\bar {\cal O}}=1$, where ${\cal O}$ is an open double cone containing 
$x_1,...,x_n$, one obtains PGI (\ref{PGI}). It has even been shown that 
(\ref{localcurrconserv}) is necessary for PGI
provided the ghost number is conserved (Sect.~4.5.2 of \cite{BD}, related ideas 
are given in \cite{Hurth}). Motivated by these facts we proceed as follows. 
In (non-perturbative) classical field theory we show that BRS-invariance of the 
Lagrangian (for constant coupling) implies local conservation of the BRS-current.
The latter holds also for the perturbative expansion of the classical fields,
i.e.~for the retarded product of classical field theory $R^{\rm class}$. Since
$R^{\rm class}$ agrees with the contribution of the tree diagrams
$R^{\rm tree}$ to the retarded product of QFT \cite{DF1,DF2}, 
we obtain the translation of (\ref{localcurrconserv}) into $R^{N\,{\rm tree}}$,
after a finite renormalization $R\rightarrow R^N$ (of tree diagrams). 
Then, proceeding analogously 
to the step from  (\ref{localcurrconserv}) to PGI (see Sect.~5.2 of \cite{DF2}),
we obtain PGI-tree for $R^{N\,{\rm tree}}$. (Up to third 
order this result is derived also in Appendix B of \cite{DF2}.)
Finally we show that PGI-tree is maintained in the transition to the corresponding time 
ordered product $T^N$ (by using results about the counting of powers of $\hbar$ 
given in Sect.~5 of \cite{DF1}).

The paper is organized as follows.
In Sect.~2 we assume that the {\it classical} BRS-current is locally conserved  
(\ref{dj(g)}), and give a model independent proof that this implies PGI-tree.
In Sect.~3 we trace back this assumption to BRS-invariance of the 
Lagrangian for constant coupling, still independently of the model.
In doing so we obtain explicit formulas for the BRS-current $j^\mu$
and the $Q$-vertex ${\cal L}_1^{\nu}$.

We then illustrate the formalism:  for
massless and massive Yang-Mills theories 
we find that our formulas for $j^\mu$ and ${\cal L}_1^{\nu}$ yield results 
which agree with the literature (Sect.~4.1 and 4.2).

Then, we turn to the main objective of this paper: massless spin-2 gauge fields 
(Sect.~4.3). From Kugo and Ojima 
\cite{KO} we recall the BRS-invariance of the Lagrangian of gravity
and show that it fits in our formalism. 
This completes our proof of PGI-tree to all orders for massless 
spin-2 gauge fields. We verify that our formula for $j^{(0)\,\mu}$ 
agrees with the literature also in the spin-2 case. 
\section{From classical current conservation to perturbative gauge invariance
for tree diagrams}\setcounter{equation}{0}
Let ${\cal P}$ be the polynomial algebra generated by the basic classical (off-shell)
fields and their partial derivatives, see Appendix A. In this Sect.~we assume that there are given
\begin{itemize}
\item an action $S_{\rm total}(g)=S_0+S(g)$ with free part $S_0=\hbar^{-1}\,\int dx\,
{\cal L}^{(0)}(x)$ and interacting part 
\beq
S(g)=\int dx\,{\cal L}(g)(x)\ ,\quad {\cal L}(g)(x):=\hbar^{-1}\>
\sum_{k=1}^\infty \kappa^k\,(g(x))^k\, {\cal L}^{(k)}(x)\ ,\label{L(g)}
\eeq
${\cal L}^{(k)}\in {\cal P}\>(\forall k=0,1,2,3,...)$, where $\kappa$ is the 
coupling constant and $g\in{\cal D}(\RR^4)$ is a test function which
switches the interaction;
\item a BRS-current 
\beq
j^{\mu}(g)(x):=\sum_{k=0}^\infty \kappa^k\,(g(x))^k\, j^{(k)\,\mu}(x)\ ,\quad
 j^{(k)\,\mu}\in {\cal P}\ ;\label{j(g)}
\eeq
\item and a  $Q$-vertex\footnote{In the terminology of \cite{S-wiley} the defining 
property of a $Q$-vertex ${\cal L}_1^{(1)\,\nu}$ is (\ref{PGI}) for $n=1$.
The fact that we use the word ``$Q$-vertex'' does not mean that we assume this identity 
to hold, it will be part of our conclusion. In addition, in our terminology a  $Q$-vertex
contains also terms of higher orders in $\kappa$.}
\beq
{\cal L}_1^{\nu}(g)(x):=\sum_{k=1}^\infty \kappa^k\,(g(x))^{(k-1)}\,
{\cal L}_1^{(k)\,\nu}(x)\ ,\quad {\cal L}_1^{(k)\,\nu}\in {\cal P}\ .\label{L_1(g)}
\eeq
\end{itemize}
Usually ${\cal L}^{(k)},\>j^{(k)\,\mu}$ and ${\cal L}_1^{(k)\,\nu}$ are of 
$(k+2)$-th order in the basic fields. The really restricting part of our assumption is
that  in {\it classical field theory} $j^{\mu}(g)$ and ${\cal L}_1^{\nu}(g)$
are related by a certain local current conservation (see (\ref{dj(g)})
below) which is a consequence of the field equations given by $S_{\rm total}(g)$.
We show that this implies PGI-tree, i.e.~the equation (\ref{PGI}) with $T^N$ replaced 
by the contribution $T^{N\,{\rm tree}}$ of its tree diagrams (on both sides of (\ref{PGI})).

We use the formalism of \cite{DF3} and \cite{DF2}, see Appendix A. 
With that a perturbative classical field (\ref{A^ret}) 
agrees exactly with the contribution of the tree diagrams
$A^\mathrm{tree}_{S(g)}(x)$ to the corresponding field $A_{S(g)}(x)$ (\ref{intfield:QFT})
of perturbative QFT: due to $A\sim\hbar^0,\> S(g)\sim\hbar^{-1}$ and
(\ref{R-tree}) it holds \cite{DF1}
\beq
A_{S(g)}(x)=A^\mathrm{tree}_{S(g)}(x)+{\cal O}(\hbar)\quad{\rm and}\quad
A^\mathrm{tree}_{S(g)}(x)\sim\hbar^0\ ,
\eeq
and hence
\beq
A^\mathrm{tree}_{S(g)}(x)\vert_{{\cal C}_{S_0}}\equiv
R_{S_0}^\mathrm{tree}\Bigl(e_\otimes^{S(g)},A(x)\Bigr)=
R_{S_0}^\mathrm{class}\Bigl(e_\otimes^{S(g)},A(x)\Bigr)\ .\label{tree=class}
\eeq
Due to this identity the classical factorization of composite fields holds also 
for  $A^\mathrm{tree}_{S(g)}(x)$ \cite{DF2},
\beq
(AB)^\mathrm{tree}_{S(g)}(x)=A^\mathrm{tree}_{S(g)}(x)\cdot B^\mathrm{tree}_{S(g)}(x)\ ,
\quad A,B\in{\cal P}\ ,\label{factorization}
\eeq
and the fields $\varphi^\mathrm{tree}_{S(g)}(x)\vert_{{\cal C}_{S_0}}$
(where $\varphi$ runs through the basic fields) satisfy the classical field equations,
which form a {\it closed} system of partial differential equations. The product on the right
side of (\ref{factorization}) is the classical product (see Appendix A), not the 
$\star$-product (\ref{*-product}). The latter is related to
the Poisson bracket (of classical field theory) by 
\beq
\{F,G\}=\lim_{\hbar\to 0}\frac{i}{\hbar}(F\star G-G\star F)\ ,\quad F,G\in {\cal F}\ .\label{Pb}
\eeq

Next we give some preparations concerning the BRS-structure.
In a BRS-model \cite{BRS} the basic fields are
\begin{itemize}
\item bosonic gauge fields with spin-1 ($A^\mu$) or spin-2 ($h^{\mu\nu}$),
\item for each gauge field there is precisely one pair of fermionic ghost fields $(u,\tilde u)$,
\item for each {\it massive} gauge field there is precisely one bosonic ghost field $\phi$,
\item in non-Abelian massive spin-1 gauge theories there is at least one 
physical Higgs field $H$
\item and there may be fermionic spinor fields.
\end{itemize}
The field algebras ${\cal F},\>{\cal F}^{(m)}_0\equiv\frac{\cal F}{{\cal J}^{(m)}}$ 
(classical product, see Appendix A) and ${\cal A}^{(m)}_0\equiv 
({\cal F}^{(m)}_0,\star_m)$ are
$\ZZ_2$-graded by the number of ghost fields: ${\cal F}={\cal F}_{\rm even}\oplus 
{\cal F}_{\rm odd}$. The ghost number of the action is even, and it is odd for the
BRS-current and the $Q$-vertex.

We assume that the free BRS-transformation $s_0$ acts linearly on the basic fields,
i.e.~$s_0\varphi$ ($\varphi$ a basic field) is a linear combination of partial derivatives of 
basic fields. This implies that $j^{(0)\,\mu}$ is quadratic in the (derivated) basic fields, since 
${\cal L}^{(0)}$ is quadratic in the (derivated) basic fields. 

We shall need some basic properties of the free BRS-charge $Q$ (or more precisely of
$d_Q$ (\ref{d_Q})). Formally, $Q$ is given by 
\beq
Q:= \int_{x^0={\rm const.}} d^3x\> j_{S_0}^{(0)\,0}(x^0,\vec{x})\ ,\label{Q}
\eeq
which is a functional on ${\cal C}_{S_0}$.
The problem with this formula is that a priori $j^{(0)\,0}_{S_0}(x^0,\vec{x})$ 
can only be integrated out in $\vec{x}$ {\it and} $x^0$ and only with a test function.  
To give a rigorous definition we follow Sect. 5.1 of \cite{DF} where a 
method of Requardt \cite{R} is used. Let $k(x^0)\,h(\vec{x})\in {\cal D}(\RR^4)$,
where $\int dx^0\,k(x^0)=1$ and $h$ is a smeared characteristic function of
$\{\vec{x}\in \RR^3, |\vec{x}|\leq R\}$ for some $R>0$. We scale the test
function such that the normalization of $k$ is maintained,
\beq
k_\lambda (x^0)\=d\lambda\, k(\lambda x^0)\ ,\quad h_\lambda 
(\vec{x})\=d h(\lambda \vec{x})\ ,
\eeq
and want to define $Q$ as the limit
\beq
Q\=d\lim_{\lambda\to 0} Q_\lambda\ ,\quad 
Q_\lambda\=d \int d^4x\,k_\lambda (x^0)\,h_\lambda (\vec{x}) \,j^{(0)\,0}_{S_0}(x^0,\vec{x})\ .
\label{Q_lambda}
\eeq
As far as we know the existence of this limit cannot be shown generally, structural information
about the concrete model is needed. However, in this paper we are not interested in
$Q$ itself, we only study the operator $d_Q:{\cal A}_0^{(m)}
\rightarrow {\cal A}_0^{(m)}$, which is given by the graded commutator\footnote{Note
that the ghost number of $Q$ is odd.}
\beq
d_Q\>F_{S_0}\=d  \lim_{\lambda\to 0}\>(Q_\lambda\star F_{S_0}\mp F_{S_0}\star Q_\lambda)
=: \lim_{\lambda\to 0}\>[Q_\lambda, F_{S_0}]^\mp_\star\ ,\label{d_Q}
\eeq
where the minus sign appears for $F\in {\cal F}_{\rm even}$ and the plus sign for
$F\in {\cal F}_{\rm odd}$. It immediately follows that $d_Q$ is a graded 
derivation\footnote{A graded derivation $D$ of a $\ZZ_2$-graded
algebra ${\cal A}$ is defined to be a {\bf linear} map $D:{\cal A}\rightarrow {\cal A}$ with
\beq
D(A\cdot  B)=D(A)\cdot B +(-1)^{\epsilon (A)}\,A\cdot D(B)\ ,
\eeq
where $A$ is of definite degree $\epsilon(A)\in\{0,1\}$.}
with respect to the $\star$-product, provided the limit
\beq
d_Q\>F_{S_0}= \lim_{\lambda\to 0}\> [Q_\lambda, F_{S_0}]^\mp_\star =
\lim_{\lambda\to 0}\int dx^0\,k_\lambda (x^0)\int d^3x\,h_\lambda (\vec{x}) \,
[j^{(0)\,0}_{S_0}(x^0,\vec{x}), F_{S_0}]^\mp_\star\label{d_Q1}
\eeq
exists. This holds indeed true \cite{DF}: namely, because of 
$\supp [j^{(0)\,0}_{S_0}(x), F_{S_0}]^\mp_\star\subset (\supp F+(\bar V_+\cup \bar V_-))$ 
we may replace $h_\lambda (\vec{x})$ by $1$
for $\lambda >0$ sufficiently small and $R$ big compared with the support of $k$.
Note that $\int d^3x\,[j^{(0)\,0}_{S_0}(x^0,\vec{x}), F_{S_0}]^\mp_\star$
exists since the region of integration is bounded; and, due to current conservation,
it is independent of $x^0$. This yields
\beq
\lim_{\lambda\to 0}\> [Q_\lambda, F_{S_0}]^\mp_\star =
\int_{x^0={\rm const.}} d^3x\,[j^{(0)\,0}_{S_0}(x^0,\vec{x}), F_{S_0}]^\mp_\star\ .\label{d_Q2}
\eeq
This result holds for the terms $\sim\hbar$ of $[\,\cdot\,,\,\cdot\,]^\mp_\star$ separately. Hence,
we may define
\beq
\{Q, F_{S_0}\}:\=d \lim_{\lambda\to 0}\> \{Q_\lambda, F_{S_0}\}
=\lim_{\hbar\to 0}\frac{i}{\hbar}\>d_Q\>F_{S_0}\label{Pb:Q}
\eeq
and obtain $\{Q, F_{S_0}\}=\int d^3x\,\{j^{(0)\,0}_{S_0}(x^0,\vec{x}), F_{S_0}\}$.

For the concrete models studied in Sect.~4 it holds 
\beq
d_Q^2=0\label{nilpotent}
\eeq
and
\beq
\omega_0(d_Q\> F_{S_0})=0\ ,\quad\forall F\in {\cal F}\ ,
\label{Q-vaccum}
\eeq
as it is verified e.g. in \cite{S-wiley} (in a Krein-Fock space representation). In this paper we 
assume only the validity of (\ref{Q-vaccum}), the nil-potency will not be needed.
(Usually, the fields are represented on an inner product space such that 
$\langle F^*\Phi,\Psi\rangle =\langle \Phi,F\Psi\rangle$ (where $\langle\>\cdot\>,\>\cdot\>
\rangle$ must be indefinite) and in that representation one proves that $Q$ 
is a nilpotent and symmetric operator which annihilates the vacuum; these properties
imply (\ref{nilpotent}) and (\ref{Q-vaccum}).)

We are now going to show that $d_Q$ is a graded derivation 
also with respect to the {\bf classical} product. This follows from 
the observation that in $d_Q\, F_{S_0}$ (\ref{d_Q}) solely the terms $\sim\hbar$ of
the $\star$-product contribute. In detail:
\begin{lemma} 
  \begin{enumerate}\item
\beq
d_Q\>F_{S_0}=-i\hbar\>\{Q,F_{S_0}\}\ ,\quad\forall F\in{\cal F}\ ,
\eeq
where $\{Q,\cdot\}$ is defined by (\ref{Pb:Q}).\\
\item
\beq
d_Q\>(F_{S_0}\,G_{S_0})=d_Q(F_{S_0})\>G_{S_0}\pm F_{S_0}\>d_Q(G_{S_0})\ ,
\eeq
where the $+$-sign holds for $F\in {\cal F}_{\rm even}$ and the $-$-sign holds 
for $F\in {\cal F}_{\rm odd}$.
 \end{enumerate}
\end{lemma}
\begin{proof} (ii) is a consequence of (i): since the graded Poisson bracket satisfies the 
graded Leibniz rule, the map $\{Q,\cdot\}$ (\ref{Pb:Q}) is a graded derivation with respect 
to the classical product.

To concentrate on the essential steps of the proof of (i) we replace $F_{S_0}$
by $\phi_1...\phi_n$ (classical product), where $\phi_j$ is (a derivative of) a basic Bose 
field $\varphi_{i_j}(x_j)$: $\phi_j=\d^{a_j}
\varphi_{i_j}(x_j)_{S_0}$. All non-vanishing terms in $[Q_\lambda,\phi_j]_\star$ have one 
contraction, hence
\beq
d_Q\>\phi_j=-i\hbar\>\{Q,\phi_j\}\ .
\eeq
To show that $d_Q(\phi_1...\phi_n)$ agrees with
\beq
-i\hbar\>\{Q,\phi_1...\phi_n\}=-i\hbar\sum_{k=1}^n\phi_1...\{Q,\phi_k\}...\phi_n\ ,
\eeq
we proceed by induction on the number $n$ of factors. By using 
\begin{itemize}
\item the recursion relation
\beq
\phi_1...\phi_{n+1}=(\phi_1...\phi_n)\star\phi_{n+1}-\sum_{l=1}^n
(\phi_1...\hat l...\phi_n)\>\omega_0(\phi_l\star\phi_{n+1})
\eeq
(which follows from the definition (\ref{*-product}) of the $\star$-product),
\item  our assumption that $s_0$ acts linearly on the basic fields which implies that 
$\{Q,\phi_j\}$ is a linear combination of derivated basic fields, 
\item and
\beq
\omega_0(\{Q,\phi_l\}\star\phi_{n+1})+\omega_0(\phi_l\star\{Q,\phi_{n+1}\})
=\frac{i}{\hbar}\> \omega_0([Q\,,\,\phi_l\star\phi_{n+1}]_\star)=0
\eeq
(where (\ref{Q-vaccum}) is used), 
\end{itemize}
one verifies the inductive step straightforwardly.
\end{proof} 
An alternative proof of the statement (ii) can be found in Lemma 3.1.1. of 
\cite{S-wiley}.

In this Sect.~we {\bf assume} that there exist a BRS-current $j^\mu(g)$ (\ref{j(g)}), 
an action $S_{\rm total}(g)=S_0+S(g)$  (\ref{L(g)}) and a $Q$-vertex
${\cal L}_1^{\nu}(g)$ (\ref{L_1(g)}) such that they fulfill the local current conservation
\beq
\d_\mu\, j^\mu(g)(x)_{S_{\rm total}(g)}=(\d_\nu g)(x)\>{\cal L}_1^\nu(g)(x)_{S_{\rm total}(g)}\ ,
\label{dj(g)}
\eeq
in classical field theory. This identity implies that the {\it BRS-current} 
$j^\mu(g)(x)_{S_{\rm total}(g)}$ {\it is conserved for} $x\in {\cal U}$
(where ${\cal U}\subset \RR^4$ is an open set)
{\it if} $g\vert_{{\cal U}}$ {\it is constant}. We will verify the assumption (\ref{dj(g)})
for concrete models in the following Sects.. 
The current conservation (\ref{dj(g)}) holds also for the corresponding 
retarded fields (\ref{A^ret:def})
and, hence, also for the perturbative expansion (\ref{A^ret}) of the latter.
With that and due to the identity (\ref{tree=class}) we obtain a statement for the 
tree diagrams of perturbative QFT:
\beq
- R^{\mathrm{tree}}_{S_0}\Bigl(e_\otimes^{S(g)},\int j(g)^{\mu}\,\d_\mu f\Bigr)=
R^{\mathrm{tree}}_{S_0}\Bigl(e_\otimes^{S(g)},\int {\cal L}_1^{\nu}(g)\,f\,\d_\nu g\Bigr)
\ .\label{curcons:R}
\eeq

We are now going to absorb the higher order terms of $S(g)$, $j(g)^{\mu}$ and
${\cal L}_1^\nu (g)$ in an admissible
renormalization of the retarded product by using the 'Main Theorem of Perturbative 
Renormalization' (Sect.~4.2 of \cite{DF3}).  We assume that ${\cal L}^{(1)}$,
$j^{(0)\,\mu}$ and ${\cal L}_1^{(1)\,\nu}$ are linearly independent fields; or that
${\cal L}^{(1)}$ and $j^{(0)\,\mu}$ are linearly independent and ${\cal L}_1^{(1)\,\nu}=0$. 
(This assumption seems to hold true in all cases of interest, see Sect.~4.) 
With that we set
\begin{gather}
\sum_{n=0}^\infty\frac{1}{n!}\,D_n\Bigl((\int g\,{\cal L}^{(1)})^{\otimes n}\Bigr)\equiv
D\Bigl(e_{\otimes}^{\int g\,{\cal L}^{(1)}}\Bigr)\=d S(g)\ ,\notag\\
D\Bigl(e_{\otimes}^{\int g\,{\cal L}^{(1)}}{\otimes} \int h_\mu\,j^{(0)\,\mu}\Bigr)\=d
\int dx\,h_\mu(x)\,j^\mu(g)(x)\ ,\notag\\
D\Bigl(e_{\otimes}^{\int g\,{\cal L}^{(1)}}{\otimes} \int h_\nu\,{\cal L}_1^{(1)\,\nu}\Bigr)\=d
\int dx\,h_\nu(x)\,{\cal L}_1^\nu(g)(x)
\end{gather}
and extend $D_n$ to a {\it linear} and {\it symmetrical} map
\beq
D_n:{\cal F}_{\rm loc}^{\otimes n}\longrightarrow {\cal F}_{\rm loc}\label{D}
\eeq
with
\begin{itemize}
\item $D_0(1)=0$, $\quad D_1(F)=F$\ ,
\item 
       \begin{equation}
       \supp\frac{\delta\,D_n(F_1\otimes\ldots\otimes F_n)}{\delta\varphi}\subset
       \bigcap_{i=1}^n \supp\frac{\delta F_{i}}{\delta\varphi}\ ,
        \end{equation}
\item Poincar\'e covariance and
\item $D(F^{\otimes n})^{*}=D((F^{*})^{\otimes n})$.
\end{itemize}
Part (iv) of the Main Theorem then states that\footnote{This formula has to be understood 
in the sense of formal power series in $\lambda$.} 
   \begin{equation}
        R^N(e_{\otimes}^{\lambda G}, F) \=d 
        R(e_{\otimes}^{D(e_{\otimes}^{\lambda G})},
        D(e_{\otimes}^{\lambda G}\otimes F)) \label{ren:intf}
   \end{equation}
defines a new retarded product $R^N$, i.e.~$R^N$ fulfills the basic properties 
(\ref{ret-prod:QFT})-(\ref{GLZ}). And, if $R$ is Poincar\'e covariant and unitary
(i.e. $R(F^{\otimes n})^*=R((F^*)^{\otimes n})$), then these properties hold also for
$R^N$. Usually $\omega_0(R_{n-1,1}(...))$ satisfies an upper bound on the scaling 
behaviour in the UV-region. The map $D$ (\ref{D}) can be chosen such that also this bound 
is maintained in the renormalization $R\longrightarrow R^N$. In terms of $R^N$
the current conservation (\ref{curcons:R}) reads
\beq
- R^N_{S_0}\Bigl(e_\otimes^{\int g\,{\cal L}^{(1)}},\int j^{(0)\,\mu}\,\d_\mu f\Bigr)=
R^N_{S_0}\Bigl(e_\otimes^{\int g\,{\cal L}^{(1)}},\int {\cal L}_1^{(1)\,\nu}\,f\,\d_\nu g\Bigr)
+{\cal O}(\hbar)\ ,\label{curcons:R^N}
\eeq
where we have also used (\ref{R-tree}). This identity implies
\begin{gather}
- R^N_{S_0}\Bigl(e_\otimes^{\int g\,{\cal L}^{(1)}}\otimes \int g\,{\cal L}^{(1)}
,\int j^{(0)\,\mu}\,\d_\mu f\Bigr)=
R^N_{S_0}\Bigl(e_\otimes^{\int g\,{\cal L}^{(1)}},\int {\cal L}_1^{(1)\,\nu}\,f\,\d_\nu g\Bigr)\notag\\
+R^N_{S_0}\Bigl(e_\otimes^{\int g\,{\cal L}^{(1)}}\otimes \int g\,{\cal L}^{(1)}
,\int {\cal L}_1^{(1)\,\nu}\,f\,\d_\nu g\Bigr)+{\cal O}(\hbar)\ .\label{curcons:R^N1}
\end{gather}

We are now going to derive PGI for $R^{N\,{\rm tree}}$.
For a given $g\in {\cal D}(\RR^4)$ let ${\cal O}$ be an open double cone such that 
$\supp g\subset {\cal O}$. Furthermore we choose $f\in {\cal D}(\RR^4)$ with
$f\equiv 1$ on a neighborhood of $\overline{\cal O}$. We decompose 
$\d_\mu f =b_\mu -a_\mu$ such that
$\supp b_\mu\cap ({\cal O}+\bar V_+)=\emptyset$ and 
$\supp a_\mu\cap ({\cal O}+\bar V_-)=\emptyset$. In the following calculation (which is mainly
taken from Sect.~5.2 of \cite{DF2}) we take into account ${\cal L}^{(1)}\sim\hbar^{-1}$,
${\cal L}_1^{(1)}\sim\hbar^{0}$ and 
\beq
\{Q,F_{S_0}\}=\{\int dx\, j_{S_0}^{(0)\,\mu}(x)\,b_\mu(x),F_{S_0}\}\ ,\quad\forall
F\in {\cal F}({\cal O})\ ,
\eeq
which follows from the locality of the Poisson bracket (i.e. $\{G_{S_0},H_{S_0}\}=0$
if the supports of $G_{S_0}$ and $H_{S_0}$ are space-like separated), the current 
conservation $\d_\mu j_{S_0}^{(0)\,\mu}=0$ and the definition of $\{Q,\cdot\}$ 
(\ref{Pb:Q}). (A detailed explanation of the same conclusion 
for the commutator is given in Appendix B of \cite{DF}.) In addition we use 
Lemma 1(i), the characterization (\ref{R-tree}) of  $R^{N\,\rm tree}$, 
$R_{0,1}(F)=F$, causality (\ref{caus}) and the GLZ relation (\ref{GLZ}):
\begin{gather}
d_Q\,R_{S_0}^{N\,{\rm tree}}\Bigl(e_\otimes^{\int g\,{\cal L}^{(1)}},\int g\,{\cal L}^{(1)}
\Bigr)
=-i\hbar\Bigl\{\int j_{S_0}^{(0)\,\mu}\,b_\mu\,,\,R_{S_0}^{N\,{\rm tree}}\Bigl(...\Bigr)\Bigr\}\notag\\
=\Bigl[\int j_{S_0}^{(0)}\,b\,,\,R_{S_0}^N\Bigl(e_\otimes^{\int g\,{\cal L}^{(1)}},
\int g\,{\cal L}^{(1)}\Bigr)\Bigr]_\star\vert_{\hbar^0}\notag\\
=\Bigl[R_{S_0}^N\Bigl(e_\otimes^{\int g\,{\cal L}^{(1)}},\int j^{(0)}\,b\Bigr)\,,\,
R_{S_0}^N\Bigl(e_\otimes^{\int g\,{\cal L}^{(1)}},
\int g\,{\cal L}^{(1)}\Bigr)\Bigr]_\star\vert_{\hbar^0}\notag\\
=i\Bigl(R_{S_0}^N\Bigl(e_\otimes^{\int g\,{\cal L}^{(1)}}\otimes\int\,j^{(0)}\,b,
\int g\,{\cal L}^{(1)}\Bigr)-R_{S_0}^N\Bigl(e_\otimes^{\int g\,{\cal L}^{(1)}}\otimes
\int g\,{\cal L}^{(1)},\int j^{(0)}\,b\Bigr)\Bigr)\vert_{\hbar^0}\ ,\label{d_QR1}
\end{gather}
where $...\vert_{\hbar^0}$ signifies that we only mean the contribution of the terms 
$\sim\hbar^0$ (which is in all terms 
of (\ref{d_QR1})-(\ref{d_QR3}) the contribution with the lowest power of $\hbar$).
Due to the the support property of $R^N$ the last retarded product vanishes and in 
the second last we may replace $b_\mu$ by $\d_\mu f$. By using the GLZ relation 
again we obtain a form to which we can apply the classical current conservation 
(\ref{curcons:R^N}) and (\ref{curcons:R^N1}):
\begin{gather}
=i\,R_{S_0}^N\Bigl(e_\otimes^{\int g\,{\cal L}^{(1)}}\otimes\int\,j^{(0)\,\mu}\,\d_\mu f,
\int g\,{\cal L}^{(1)}\Bigr)\vert_{\hbar^0}\notag\\
=\Bigl[R_{S_0}^N\Bigl(e_\otimes^{\int g\,{\cal L}^{(1)}},\int j^{(0)}\,\d f\Bigr)\,,\,
R_{S_0}^N\Bigl(e_\otimes^{\int g\,{\cal L}^{(1)}},
\int g\,{\cal L}^{(1)}\Bigr)\Bigr]_\star\vert_{\hbar^0}\notag\\
+i\,R_{S_0}^N\Bigl(e_\otimes^{\int g\,{\cal L}^{(1)}}\otimes
\int g\,{\cal L}^{(1)},\int j^{(0)}\,\d f\Bigr)\vert_{\hbar^0}\notag\\
=-\Bigl[R^N_{S_0}\Bigl(e_\otimes^{\int g\,{\cal L}^{(1)}},\int {\cal L}_1^{(1)}\,\d g\Bigr)
\,,\,R_{S_0}^N\Bigl(e_\otimes^{\int g\,{\cal L}^{(1)}},
\int g\,{\cal L}^{(1)}\Bigr)\Bigr]_\star\vert_{\hbar^0}\notag\\
-i\,R^N_{S_0}\Bigl(e_\otimes^{\int g\,{\cal L}^{(1)}}\otimes \int g\,{\cal L}^{(1)}
,\int {\cal L}_1^{(1)}\,\d g\Bigr)\vert_{\hbar^0}
-i\,R^N_{S_0}\Bigl(e_\otimes^{\int g\,{\cal L}^{(1)}},\int {\cal L}_1^{(1)}\,\d g\Bigr)
\vert_{\hbar^0}\ ,\label{d_QR2}
\end{gather}
where $f(x)\,\d_\nu g(x) =\d_\nu g(x)$ is taken into account. By 
means of the GLZ relation
the first three retarded products can be expressed by one retarded product.
Applying (\ref{R-tree}) again we end up with
\begin{gather}
=-i\,R^{N\,{\rm tree}}_{S_0}\Bigl(e_\otimes^{\int g\,{\cal L}^{(1)}}\otimes
\int {\cal L}_1^{(1)}\,\d g ,\int g\,{\cal L}^{(1)}\Bigr)
-i\,R^{N\,{\rm tree}}_{S_0}\Bigl(e_\otimes^{\int g\,{\cal L}^{(1)}},\int {\cal L}_1^{(1)}\,
\d g\Bigr)\ .\label{d_QR3}
\end{gather}
(\ref{d_QR1})=(\ref{d_QR3}) is PGI-tree for $R^N$.

Next we show that PGI-tree is maintained in the transition to the corresponding 
{\it connected} time ordered product $T^N_c$. Following
Sect.~5.2 of \cite{DF1} we define recursively the {\it connected product}
\begin{equation}
(F_1\star...\star F_n)_c\=d (F_1\star...
\star F_n)-\sum_{|P|\geq 2}\prod_{J\in P}(F_{j_1}\star...\star 
F_{j_{|J|}})_c\ ,\label{conn}
\end{equation}
where $\{j_1,...,j_{|J|}\}=J$, $j_1<...<j_{|J|}$,  the sum runs over 
all partitions $P$ of $\{1,...,n\}$ in at least two subsets and
$\prod$ means the classical product. Identifying the vertices within each
$F_j$, $(F_1\star...\star F_n)_c$ is precisely the contribution of all connected 
diagrams to $F_1\star...\star F_n$. According to Proposition 1 of \cite{DF2} it holds
\beq
(F_1\star...\star F_n)_c={\cal O}(\hbar^{n-1})\quad{\rm if}\quad
F_1,...,F_n\sim\hbar^0\ .
\eeq
Since solely connected diagrams contribute to the retarded products $R_{n,1}$,
we conclude from formula (E.6) in \cite{DF3} that the connected part $T^N_c$ 
of $T^N$ is obtained from $R^N$ by
\begin{gather}
  T^N_{n\,c}(F^{\otimes n})=\sum_{k=1}^n i^{k-n}\sum_{l_1+...+l_k=n-k}
N(n,k,l_1,...,l_k)\notag\\
\cdot\bigl(R^N_{l_1,1}(F^{\otimes l_1},F)\star...\star 
R^N_{l_k,1}(F^{\otimes l_k},F)\bigr)_c\ ,
\end{gather}
where $N(n,k,l_1,...,l_k)\in\RR$ is a combinatorical factor.
We find $T^N_{n\,c}(F^{\otimes n})= {\cal O}(\hbar^{n-1})$ if $F\sim\hbar^0$.
The contribution of the tree diagrams is that part with the lowest power of $\hbar$:
\beq
T^{N\,{\rm tree}}_{n\,c}(F^{\otimes n})=\sum...\bigl(R^{N\,{\rm tree}}_{l_1,1}(...)\star...\star 
R^{N\,{\rm tree}}_{l_k,1}(...)\bigr)_c\vert_{\hbar^{n-1}}\ .
\eeq
Because $d_Q$ is a graded derivation with respect to the classical and the $\star$-product,
it is also a graded derivation with respect to the {\it connected} product (\ref{conn}).
Hence, PGI for $R^{N\,{\rm tree}}$ implies
\begin{gather}
d_Q\,T^{N\,{\rm tree}}_{n\,c\>S_0}\Bigl((\int g\,{\cal L}^{(1)})^{\otimes n}\Bigr)=\sum...
\frac{d}{d\lambda}\vert_{\lambda =0}\Bigl(R^{N\,{\rm tree}}_{l_1,1\>S_0}\Bigl(
(\int g\,{\cal L}^{(1)}-i\lambda\,\int {\cal L}^{(1)}_1\,\d g)^{\otimes (l_1+1)}\Bigr)\star...\notag\\
\star R^{N\,{\rm tree}}_{l_k,1\>S_0}\Bigl(
(\int g\,{\cal L}^{(1)}-i\lambda\,\int {\cal L}^{(1)}_1\,\d g)^{\otimes (l_k+1)}\Bigr)
\Bigr)_c\vert_{\hbar^0}\notag\\
=-i\,n\>T^{N\,{\rm tree}}_{n\,c\>S_0}\Bigl(\int {\cal L}^{(1)}_1\,\d g\otimes 
(\int g\,{\cal L}^{(1)})^{\otimes (n-1)}\Bigr)\ .\label{gauinv:T_c}
\end{gather}

Finally PGI remains valid also in the step from 
$T^{N\,{\rm tree}}_{c}$ to $T^{N\,{\rm tree}}$, because each tree diagram is the 
{\it classical} product of its connected components (the latter are also tree diagrams)
and since $d_Q$ is a graded derivation with respect to the classical product.
In detail,
\begin{equation}
T^{N\,{\rm tree}}_n(F_1\star...\star F_n)= T^{N\,{\rm tree}}_{n\,c}(F_1\star...
\star F_n)+\sum_{|P|\geq 2}\prod_{J\in P}T^{N\,{\rm tree}}_{|J|\, c}(F_{j_1}\star...\star 
F_{j_{|J|}}) ,\label{T-tree}
\end{equation}
where $\prod$, $P$ and $J$ are as in (\ref{conn}). With that the statement is obtained 
analogously to (\ref{gauinv:T_c}).
\section{From BRS-invariance of the Lagrangian to local 
conservation of the classical BRS-current}\setcounter{equation}{0}
The proof in the preceding Sect.~is based on the local BRS-current conservation 
(\ref{dj(g)}) for {\bf classical field theory}. In this Sect.~we show that this 
assumption follows from {\it BRS-invariance of the Lagrangian for constant coupling}. 
The latter is verified for concrete models in Sect.~4, in particular for massless
spin-2 gauge fields.

In this procedure, solely the Lagrangian and the BRS-transformation $s$ for constant 
coupling $\kappa$ need to be given. With that we construct a BRS-transformation 
$s(g)$ for local coupling $\kappa\,g(x)$ and a corresponding local
Noether current $j^\mu(g)$. We show that the divergence of this local BRS-current
is indeed of the form  (\ref{dj(g)}) and in doing so we obtain an explicit formula
for the $Q$-vertex ${\cal L}^\mu_1(g)$. 

The Lagrangian and the BRS-transformation are assumed to be {\it formal power 
series in} $\kappa$ and we understand all equations in the sense of formal power series.
However, we point out that Sects.~3 and 4 are {\bf non-perturbative} classical field theory.
\medskip

We assume that a BRS-invariant Lagrangian
\beq
{\cal L}_{\rm total}=\sum_{n=0}^\infty\kappa^n\,{\cal L}^{(n)}
=:{\cal L}^{(0)}+{\cal L}_{\rm int}\ ,\quad
{\cal L}^{(n)}\in {\cal P}\ ,\label{L_total}
\eeq
is given, where the free part ${\cal L}^{(0)}$ is quadratic in the (derivated) basic fields
and higher than first derivatives of the basic fields do not appear in each ${\cal L}^{(n)}$. 
By BRS-invariance we mean that there exists
\beq 
I^\mu=\sum_{n=0}^\infty \kappa^n\, I^{(n)\,\mu}\ ,\quad
I^{(n)\,\mu}\in {\cal P}\ ,\label{I}
\eeq
such that
\beq
s\,{\cal L}_{\rm total}=-\d_\mu\,I^\mu\label{sL1}
\eeq
without using the field equations. We admit BRS-transformations which are formal power 
series in $\kappa$:
\beq 
s=\sum_{n=0}^\infty \kappa^n\, s_n\ .\label{s}
\eeq
We assume that $s$ is given on the basic fields $\varphi$ and that it is extended to a
linear map $s:{\cal P}\rightarrow {\cal P}$ (more precisely, to a formal power series of 
linear maps $s_n:{\cal P}\rightarrow {\cal P}$) by setting
\beq
s\,(\d^a\varphi)\=d\d^a(s\,\varphi)\ ,\quad a\in\NN_0^4\ ,\label{sdphi}
\eeq
for all basic fields $\varphi$, and by requiring that $s$ is a graded derivation.
It follows 
\beq
s\,(\d^\mu \,A)=\d^\mu\,(s\,A)\ ,\quad\forall A\in {\cal P}\ ,\label{[s,d]}
\eeq
and that each $s_n$ is a graded derivation which commutes with partial derivatives
(\ref{[s,d]}). In order that the BRS-symmetry (\ref{sL1}) can be used to construct the 
corresponding quantum gauge theory it is needed that $s$ is nilpotent modulo the 
field equations:
\beq
s^2\,(A)\>\vert_{{\cal C}_{S_{\rm total}}}=0\ ,\quad\forall A\in{\cal P}\ ,\label{s^2=0}
\eeq
where $S_{\rm total}=\int dx\, {\cal L}_{\rm total}$. However, in the proof of 
PGI-tree given in this paper the nil-potency of $s$ is not used.
\medskip

We first recall the construction of the Noether current for {\bf constant coupling} $\kappa$
(cf.~any book on classical field theory). By using the derivation property of $s$, 
(\ref{[s,d]}) and the field equations, we get
\beq
(s\,{\cal L}_{\rm total})_{S_{\rm total}}=\Bigl(
\frac{\d{\cal L}_{\rm total}}{\d\varphi_i}\,s\varphi_i+\frac{\d{\cal L}_{\rm total}}{\d\varphi_{i,\mu}}
\,(s\varphi_i)_{,\mu}\Bigr)_{S_{\rm total}}=\d_\mu\Bigl(
\frac{\d{\cal L}_{\rm total}}{\d\varphi_{i,\mu}}\,s\varphi_i\Bigr)_{S_{\rm total}}\ ,
\label{sL2}
\eeq
where it is summed over all basic fields $\varphi_i$.
The equality of (\ref{sL1}) (restricted to ${\cal C}_{S_{\rm total}}$) and (\ref{sL2}) yields that
\beq
j^\mu\=d -\Bigl(\frac{\d{\cal L}_{\rm total}}{\d\varphi_{i,\mu}}\,s\varphi_i+I^\mu\Bigr) =
\sum_{n=0}^\infty \kappa^n\, j^{(n)\,\mu}\ ,\label{j}
\eeq
is a conserved BRS-current,
\beq
\d_\mu\,j^\mu_{S_{\rm total}}=0\ .\label{dj=0}
\eeq
For later purpose we note
\beq
 j^{(n)\,\mu}=-\Bigl(\sum_{k=0}^n\frac{\d{\cal L}^{(k)}}{\d\varphi_{i,\mu}}\>s_{n-k}\,\varphi_i+
I^{(n)\>\mu}\Bigr)\ .\label{j^n}
\eeq
\medskip

We are now going to generalize this construction of the BRS-current to 
{\bf local couplings} $\kappa\,g(x)$, $g\in {\cal D}(\RR^4)$. 
Roughly speaking we do this by replacing $\kappa$ by $\kappa\,g(x)$ everywhere.
In detail: from ${\cal L}^{(n)}$ (\ref{L_total}) and $j^{(n)\,\mu}$ (\ref{j^n})
we construct ${\cal L}(g)$ (\ref{L(g)}) and $j^{\mu}(g)$ (\ref{j(g)}).
In the Lagrangian ${\cal L}_{\rm total}$ (\ref{L_total}) we replace the interacting part
${\cal L}_{\rm int}=\sum_{\kappa=1}^\infty \kappa^n {\cal L}^{(n)}$ by ${\cal L}(g)$ 
(\ref{L(g)}) and we use the same notations $S_0,\>S(g)$ and $S_{\rm total}(g)$
as in Sect.~2.
The {\it local BRS-transformation} $s(g)$ is also a formal power series in $\kappa$,
\beq 
s(g)=\sum_{n=0}^\infty \kappa^n\, s_n(g)\ .\label{s(g)}
\eeq
We determine $s(g)$ by requiring that 
it is a graded derivation and by its action on the basic fields $\varphi$ 
and their partial derivatives:\footnote{Motivated by Sect.~5.2 of \cite{DF2} an
alternative, more explicit definition of $s(g)$ seems to be natural: namely
(\ref{s(g)}) with
\beq
s_n(g):=\int dx\,(g(x))^n\,\tilde s_n(x)\ ,\quad
\tilde s_n(x):= (s_n\varphi_i)(x)\,\frac{\delta}{\delta\varphi_i(x)}\ ,\label{s(g):alternat}
\eeq
where it is summed over all basic fields $\varphi_i$ and
$s_n$ is given from the model with constant coupling. Obviously, the so defined $s(g)$
is a graded derivation and one easily verifies that it satisfies (\ref{s(g)phi}) and
(\ref{s(g)dphi}). Hence, this definition (\ref{s(g):alternat}) agrees with the 
definition given in the main text.}
\beq
s(g)\,\varphi(x)\=d \sum_{n=0}^\infty \kappa^n\,(g(x))^n\, s_n\,\varphi(x)\ ,\label{s(g)phi}
\eeq
(where $s_n$ is given from the model with constant coupling (\ref{s})) and
\beq
s(g)\,(\d^a\varphi)\=d \d^a (s(g)\,\varphi)\ ,\quad a\in\NN_0^4\ .\label{s(g)dphi}
\eeq
As in (\ref{[s,d]}) it follows $s(g)\,(\d^\mu \,A)=\d^\mu\,(s(g)\,A)\ ,\>\forall
A\in {\cal P}$. Since $s(g)$ is a graded derivation which commutes with partial derivatives,
this holds also for
\beq
s_k(g)\=d \frac{1}{k!}\>\frac{d^k}{d\kappa^k}\vert_{\kappa =0}\, s(g)\ ,\quad
k\in\NN_0\ .
\eeq
For a basic field $\varphi$ we obtain
\begin{gather}
s_k(g)\,\varphi(x)=(g(x))^k\,s_k\,\varphi(x)\ ,\notag\\
s_k(g)\,\varphi^{,\mu}=\bigl(g^k\,s_k\,\varphi\bigr)^{,\mu}=
g^k\,s_k(\varphi^{,\mu})+g^{,\mu}\,k\,g^{(k-1)}\,s_k\,\varphi\ .\label{s_1(g)}
\end{gather}
$s(g)$ is in general not nilpotent modulo the field equations. But, if $g\vert_{\cal O}=1$
for some region ${\cal O}\subset\RR^4$, we have $s(g)\vert_{{\cal F}({\cal O})}= 
s\vert_{{\cal F}({\cal O})}$ (where $s$ is the BRS-transformation of the corresponding 
model with constant coupling) and hence 
\beq
(s(g))^2\,(F)\vert_{{\cal C}_{S_{\rm total}(g)}}=0\ ,\quad\forall F\in {\cal F}({\cal O})\ .
\eeq

As in (\ref{sL2}) the field equations for $S_{\rm total}(g)$ imply
\begin{gather}
\Bigl(s(g)\,{\cal L}_{\rm total}(g)\Bigr)_{S_{\rm total}(g)}=\d_\mu\Bigl(
\frac{\d{\cal L}_{\rm total}(g)}{\d\varphi_{i,\mu}}\>
s(g)\,\varphi_i\Bigr)_{S_{\rm total}(g)}\notag\\
=\d_\mu\sum_{n=0}^\infty \kappa^n\,g^n\,\sum_{k=0}^n\Bigl(
\frac{\d{\cal L}^{(k)}}{\d\varphi_{i,\mu}}\>s_{n-k}\,\varphi_i\Bigr)_{S_{\rm total}(g)}\ .
\label{s(g)L2}
\end{gather}
By using the derivation property of $s_l(g)$ and of $s_l$ we obtain
\begin{gather}
s_l(g)\,{\cal L}^{(k)}=\frac{\d{\cal L}^{(k)}}{\d\varphi_i}\>g^l\,s_l\,\varphi_i +
\frac{\d{\cal L}^{(k)}}{\d\varphi_{i,\mu}}\>\bigl(g^l\,s_l\,(\varphi_{i,\mu})+g_{,\mu}\,
l\,g^{(l-1)}\,s_l\,\varphi_i\bigr)\notag\\
=g^l\,s_l\,{\cal L}^{(k)}+ g_{,\mu}\,l\,g^{(l-1)}\,
\frac{\d{\cal L}^{(k)}}{\d\varphi_{i,\mu}}\>s_l\,\varphi_i\ .
\end{gather}
With that we get
\begin{gather}
s(g)\,{\cal L}_{\rm total}(g)=\sum_{n=0}^\infty \kappa^n\,\sum_{k=0}^n\,g^k\,
s_{n-k}(g)\,{\cal L}^{(k)}\notag\\
=\sum_{n=0}^\infty \kappa^n\,g^n\,\sum_{k=0}^n s_{n-k}{\cal L}^{(k)}
+g_{,\mu}\,\sum_{n=1}^\infty \kappa^n\,g^{(n-1)}\,\sum_{k=0}^{n-1} (n-k)\>
\frac{\d{\cal L}^{(k)}}{\d\varphi_{i,\mu}}\>s_{n-k}\,\varphi_i\ .\label{s(g)L1}
\end{gather}
With $s=\sum_{n=0}^\infty \kappa^n\,s_n$ and (\ref{sL1}) the first term is equal to
\begin{gather}
\sum_{n=0}^\infty \kappa^n\,g^n\,\bigl(s\,{\cal L}_{\rm total}\bigr)^{(n)}
=-\sum_{n=0}^\infty \kappa^n\,g^n\,\d_\mu\,I^{(n)\,\mu}\notag\\
=-\d_\mu\,\sum_{n=0}^\infty \kappa^n\,g^n\,I^{(n)\,\mu}+
g_{,\mu}\,\sum_{n=1}^\infty \kappa^n\,n\,g^{(n-1)}\,I^{(n)\,\mu}\ .\label{s(g)L1a}
\end{gather}
We insert (\ref{s(g)L1a}) into (\ref{s(g)L1}) and the resulting equation into the left side
of (\ref{s(g)L2}), and then we use (\ref{j^n}). This yields the local current
conservation (\ref{dj(g)}),
where ${\cal L}_1^\mu(g)$ is given by (\ref{L_1(g)}) and
\beq
{\cal L}_1^{(n)\,\mu}\=d -\Bigl(\sum_{k=0}^{n-1} (n-k)\>
\frac{\d{\cal L}^{(k)}}{\d\varphi_{i,\mu}}\>s_{n-k}\,\varphi_i
+n\>I^{(n)\,\mu}\Bigr)\ ,\quad n=1,2,...\ .\label{L_1^n}
\eeq
With that the proof of PGI-tree is complete for models with a BRS-invariant 
Lagrangian (\ref{sL1}).
\medskip

$d_Q$ can be viewed as a graded derivation $d_Q:{\cal P}\vert_{{\cal C}_{S_0}}
\rightarrow {\cal P}\vert_{{\cal C}_{S_0}}$.
Interpreting $d_Q$ in this sense we would like to prove
\beq
d_Q\>A_{S_0}=i\>(s_0\,A)_{S_0}\ ,\quad\forall A\in {\cal P}\ .\label{d_Q=s_0}
\eeq

Since $d_Q$ and $s_0$ are both graded derivations which commute with partial derivatives
it suffices to prove (\ref{d_Q=s_0}) for $A$ running through all basic fields $\varphi_i$.
In models with solely massless fields it usually holds that $s_0\,\varphi_i$ is
a divergence, i.e.
\beq
s_0\,\varphi_i=\d_\mu\,\phi^\mu_i\quad {\rm for\>some}\quad \phi^\mu_i\in {\cal P}\ ,
\label{s_0=div}
\eeq
for all basic fields $\varphi_i$. If this holds true, (\ref{d_Q=s_0}) follows from part (i) of the 
following Corollary of PGI-tree.
\begin{cor} Let a free Lagrangian ${\cal L}^{(0)}$ be given which is invariant with respect to a
free BRS-transformation $s_0$,
\beq
s_0\,{\cal L}^{(0)}=-\d_\mu\,I^{(0)\,\mu}\quad {\it for\>some}\quad I^{(0)\,\mu}\in {\cal P}\ .
\eeq
In terms of the corresponding Noether current $j^{(0)\,\mu}$ (\ref{j^n})
we define $d_Q$ by (\ref{Q_lambda})-(\ref{d_Q}). 
 \begin{enumerate}
 	\item If $s_0\,A=\d_\mu\, B^\mu$ for some $B^\mu$, then $A$ 
satisfies the relation (\ref{d_Q=s_0}).
\item For an arbitrary $P\in {\cal P}$ it holds
\beq
\d^\nu\,\bigl(d_Q\>P_{S_0}-i\>(s_0\,P)_{S_0}\bigr)=0\ .\label{d_Q=s_01}
\eeq
 \end{enumerate}
\end{cor}
\begin{proof} (i) To the given free Lagrangian we add the interaction
${\cal L}_{\rm int}:=\kappa\, A$ and choose $s:=s_0$. This model is BRS invariant:
\beq
s {\cal L}_{\rm total}=s_0\,{\cal L}^{(0)}+\kappa\,s_0\,A=-\d_\mu (I^{(0)\,\mu}-\kappa\, B^\mu)\ .
\eeq
So, our assumption (\ref{sL1}) is satisfied and, hence, PGI-tree holds with
\beq
{\cal L}_1^{(1)\,\mu}=-I^{(1)\,\mu}=B^\mu\ ,
\eeq
where (\ref{L_1^n}) is used. To first order PGI-tree reads
\beq
d_Q\>A_{S_0}=d_Q\> {\cal L}^{(1)}_{S_0}=i\>\d_\mu\,{\cal L}^{(1)\,\mu}_{1\>\>S_0}
=i\>\d_\mu\,B^\mu_{S_0}=i\>(s_0\,A)_{S_0}\ .
\eeq
(ii) Since $s_0\,(\d^\nu P)=\d_\mu\,(\eta^{\mu\nu}\>s_0\,P)$, part (i)
applies to $A=\d^\nu P$.
\end{proof}

\noindent {\it Remarks: (1)} From (\ref{d_Q=s_0}) and PGI-tree it follows
\beq
i\,(s_0\,{\cal L}^{(1)})_{S_0}=d_Q\,{\cal L}^{(1)}_{S_0}=i\,\d_\nu {\cal L}^{(1)\,\nu}_{1\>S_0}\ .
\eeq
The relation
\beq
(s_0\,{\cal L}^{(1)})_{S_0}=\d_\nu {\cal L}^{(1)\,\nu}_{1\>S_0}
\eeq
(where ${\cal L}_1^{(1)\,\mu}$ is given by 
${\cal L}_1^{(1)\,\mu}=-\frac{\d{\cal L}^{(0)}}{\d\varphi_{i,\mu}}\>s_1\,\varphi_i
-I^{(1)\,\mu}$  (\ref{L_1^n}))
can be verified directly, i.e.~without using PGI-tree. Namely,
by using $-\d_\mu\,I^{(1)\,\mu}=s_0\,{\cal L}^{(1)}+s_1\,{\cal L}^{(0)}$ (\ref{sL1}),
the derivation property of $s_1$  and the free field equations we obtain
\begin{gather}
\bigl(s_0\,{\cal L}^{(1)}-\d_\mu {\cal L}_1^{(1)\,\mu}\bigr)_{S_0}=
\Bigl(-s_1\,{\cal L}^{(0)}+\d_\mu\Bigl(\frac{\d{\cal L}^{(0)}}{\d\varphi_{i,\mu}}\>
s_1\,\varphi_i\Bigr)\Bigr)_{S_0}\notag\\
=\Bigl(-\frac{\d{\cal L}^{(0)}}{\d\varphi_{i}}\>s_1\,\varphi_i-
\frac{\d{\cal L}^{(0)}}{\d\varphi_{i,\mu}}\>(s_1\,\varphi_i)_{,\mu}+
\d_\mu\Bigl(\frac{\d{\cal L}^{(0)}}{\d\varphi_{i,\mu}}\>s_1\,\varphi_i\Bigr)\Bigr)_{S_0}\notag\\
=\Bigl(\d_\mu\frac{\d{\cal L}^{(0)}}{\d\varphi_{i,\mu}}-\frac{\d{\cal L}^{(0)}}{\d\varphi_{i}}
\Bigr)_{S_0}\>(s_1\,\varphi_i)_{S_0}=0\ .
\end{gather}
\noindent {\it (2)} If higher ($\geq 2$) derivatives of the basic fields appear in 
${\cal L}_{\rm total}$ (\ref{L_total}) the construction of a conserved BRS-current is still 
possible in the case of constant coupling: the field equations
\beq
\sum_{l\in\NN_0}(-1)^l\>\d_{\mu_1}...\d_{\mu_l}\,\frac{\d\,{\cal L}_{\rm total}}
{\d\,\varphi_{i,\mu_1...\mu_l}}=0
\eeq
imply that the current
\beq
j^\mu\=d -\Bigl(\sum_{l\in\NN_0}\sum_{j=0}^l(-1)^j\>\d_{\mu_1}...\d_{\mu_j}
\Bigl(\frac{\d\,{\cal L}_{\rm total}}{\d\,\varphi_{i,\mu\mu_1...\mu_l}}\Bigr)\> 
s\varphi_{i,\mu_{j+1}...\mu_l} + I^\mu\Bigr)
\eeq
is conserved. But in case of a local coupling terms $\sim g_{,\mu\mu_1}(x),\>\sim
g_{,\mu}(x)\, g_{,\mu_1}(x),...$ appear in (\ref{s(g)L2})-(\ref{s(g)L1}) and, hence, also in the 
divergence of the local BRS-current, i.e.~on the right side of (\ref{dj(g)}).
\section{Models}\setcounter{equation}{0}
\subsection{Massless Yang-Mills theories}
To point out the similarity of the BRS-symmetry for massless
spin-1 and spin-2 gauge fields we first recall the spin-1 case.
In terms of the covariant derivative
\beq
D^\mu_{ab}\=d \delta_{ab}\d^\mu -\kappa\,f_{abc}\,A^\mu_c\ ,
\eeq
(where $f_{abc}$ is totally antisymmetric and satisfies the Jacobi identity)
the BRS-transformation reads
\beq
s\,A_a^\mu=D^\mu_{ab}\,u_b\ ,\quad s\,u_a=-\frac{\kappa}{2}\,f_{abc}\,
u_b\,u_c\ ,\quad s\tilde u_a=-\d_\mu\,A^\mu_a\ .
\eeq
Due to
\beq
F^{\mu\nu}_{a}\equiv\d^\mu A^\nu_a-\d^\nu A^\mu_a+\kappa\, g\,f_{abc}
A^\mu_b\, A_c^\nu\label{F}
\eeq
the Yang-Mills Lagrangian
\beq
{\cal L}_{\rm YM}=-\frac{1}{4}\> F^{\mu\nu}_{a}\,F_{a\,\mu\nu}\label{L_YM}
\eeq
is of the form ${\cal L}_{\rm YM}={\cal L}_{\rm YM}^{(0)}+\kappa\,
{\cal L}_{\rm YM}^{(1)}+\kappa^2\,{\cal L}_{\rm YM}^{(2)}$.
The gauge fixing Lagrangian is of zeroth order in $\kappa$,
\beq
{\cal L}_{\rm GF}=-\frac{1}{2}\>(\d_\nu A^\nu_a)^2\ ,
\eeq
where we choose Feynman gauge.
However, the ghost Lagrangian has also a term linear in $\kappa$,
\beq
{\cal L}_{\rm ghost}=\d_\mu\tilde u_a\> s\,A^\mu_a\ .\label{L_ghost}
\eeq
Note $s_k=0,\>\forall k\geq 2$, and ${\cal L}_{\rm total}^{(j)}=0,\>\forall j\geq 3$,
where ${\cal L}_{\rm total}\=d {\cal L}_{\rm YM}+{\cal L}_{\rm GF}+{\cal L}_{\rm ghost}$.
The BRS-transformation is nilpotent modulo the field equations,\footnote{If one 
introduces the Nakanishi-Lautrup fields $B_a$ \cite{LN}, the BRS transformation 
$s$ is nilpotent in ${\cal P}$ (i.e.~without using the field equations). $s$ is then
modified as follows: $s\,\tilde u_a=-B_a$ and $s\,B_a=0$.\label{fn:NL}}
\beq
s^2\,A_a^\mu=0\ ,\quad s^2\,u_a=0\, ,\quad
s^2\tilde u_a=\frac{\delta\, S_{\rm total}}{\delta\,\tilde u_a}\ ,
\eeq
with $S_{\rm total}\=d\int dx\,{\cal L}_{\rm total}$. (Since $s$ is a graded derivation which 
commutes with partial derivatives the vanishing of $s^2$ on the basic fields implies
$s^2=0$.)

We are now going to verify the BRS-invariance of ${\cal L}_{\rm total}$.
$s\,{\cal L}_{\rm YM}$ vanishes, since the BRS-transformation of $A^\mu_a$ 
has the form of an infinitesimal gauge transformation
and since the gauge variation of ${\cal L}_{\rm YM}$
vanishes. For ${\cal L}_{\rm GF}$ and ${\cal L}_{\rm ghost}$ we obtain
\begin{gather}
s\,({\cal L}_{\rm GF}+{\cal L}_{\rm ghost})=-(\d_\nu A^\nu_a)\>\d_\mu (s\,A^\mu_a)
-(\d_\mu\d_\nu A^\nu_a)\> s\,A^\mu_a-\d_\mu\tilde u_a\>s^2A_a^\mu\notag\\
=-\d_\mu\,\bigl((\d_\nu A^\nu_a)\> D^\mu_{ab}\,u_b\bigr)=:-\d_\mu\,I^\mu\ .
\label{s(GF+ghost)}
\end{gather}

Our formula (\ref{L_1^n}) yields the following explicit expressions for the $Q$-vertex,
which agree with the literature \cite{DHKS,S-wiley,BD}:
\begin{gather}
{\cal L}_1^{(1)\,\nu}=f_{abc}[A_{a\,\mu}\, u_b\,(\d^\nu A^\mu_c-\d^\mu A^\nu_c)
-\frac{1}{2}u_a\,u_b\,\d^\nu\tilde u_c]\ , \\
{\cal L}_1^{(2)\,\nu}=f_{abr}f_{cdr}\,A_{a\,\mu}\, u_b\, A_c^\nu\, A_d^\mu\ , \\
{\cal L}_1^{(j)\,\nu}=0\ ,\quad\forall j\geq 3\ .
\end{gather}
For the zeroth order of the BRS-current our expression (\ref{j^n}) gives
\begin{gather}
j^{(0)\,\mu}=\d_\tau A^\tau_a\, \d^\mu u_a- (\d^\mu\d_\tau A^\tau_a)\, u_a\notag\\
-\d_\nu\,\bigl((A^{\mu,\nu}_a-A^{\nu,\mu}_a)\,u_a\bigr)+(\square A_a^\mu)\,u_a\ .
\end{gather}
The terms in the second line do not contribute to $d_Q$ (\ref{d_Q1})-(\ref{d_Q2}). 
This is obvious for the last term due to $\square A_{a\,{S_0}}^\mu=0$. Turning to
the second last term  we point out that generally a term of the form
$\d_\nu(K^{\mu\nu}-K^{\nu\mu})$ does not contribute to $d_Q$ (\ref{d_Q2}): 
\beq
\int_{x^0={\rm const.}} d^3x\,\d^x_l\> [(K_{S_0}^{0l}(x^0,\vec{x})
-K_{S_0}^{l0}(x^0,\vec{x})), F_{S_0}]^\mp_\star=0\ .\label{K=0}
\eeq
Hence, $d_Q$ can be constructed from the free BRS-current 
$(\d A)\, \d^\mu u- (\d^\mu\d A)\, u$ as it is done in \cite{DHKS,S-wiley,BD}.
\subsection{Massive spin-1 fields}
It is instructive to see how the BRS-formalism of the preceding Subsect.~is modified 
for {\it massive} fields. To simplify the notations we consider the most simple non-Abelian 
model: the $SU(2)$ Higgs-Kibble model, which describes three spin-1 fields,
$A_a^\mu,\>a=1,2,3$, and the bosonic scalar fields form a complex $SU(2)$ doublet,
\begin{equation}
  \Phi=\frac{1}{\sqrt{2}}\left(
  \begin{array}{c}
        \phi_2+i\phi_1  \\
       v+H-i\phi_3
  \end{array}\right)\ . \label{Phi}
\end{equation}
The shift $v\in\RR_+$ will be chosen such that the Higgs potential has a non-trivial 
minimum at $\phi =\frac{1}{\sqrt{2}}\left(
  \begin{array}{c}
        0  \\
       v \end{array}\right)$ (\ref{v}), as usual in the Higgs mechanism. The corresponding 
covariant derivative is
\beq
{\cal D}^\mu=({\bf 1}\,\d^\mu-\kappa\,\frac{i}{2}\,A^\mu_a\,\sigma_a)\ ,
\eeq
where $(\sigma_a)_{a=1,2,3}$ are the Pauli matrices. Requiring renormalizability 
and $SU(2)$-invariance the Higgs Lagrangian takes the form
\beq
{\cal L}_\Phi=({\cal D}_\mu\Phi)^+({\cal D}^\mu\Phi)+\mu^2\,\Phi^+\Phi
-\lambda\,(\Phi^+\Phi)^2\ ,\label{L_Phi}
\eeq
Choosing $\mu^2>0$ and $\lambda>0$ one obtains
\beq
v^2=\frac{\mu^2}{\lambda}\ .\label{v}
\eeq
The $SU(2)$-invariant Yang-Mills Lagrangian ${\cal L}_{\rm YM}$ is still given by 
(\ref{F})-(\ref{L_YM}). By the Higgs mechanism the three spin-1 fields $A^\mu_a\>
(a=1,2,3)$ get the same mass
\beq
m=\frac{\kappa\,v}{2}>0\ .\label{m}
\eeq

From the quantization of spin-1 fields one knows that in the massive case the condition 
$\d_\mu A^\mu_a=0$ on observables is replaced by $(\d_\mu A^\mu_a+m\phi_a)=0$.
(For free fields this is explained in Sect.~3 of \cite{DF4}.) Therefore,
\beq
s\,\tilde u_a=-(\d_\mu A^\mu_a+m\phi_a)\ ,
\eeq
and this replacement appears also in the gauge fixing term
\beq
{\cal L}_{\rm GF}=-(\d_\mu A^\mu_a+m\phi_a)^2={\cal L}_{\rm GF}^{(0)}\ ,
\eeq
where we choose again Feynman gauge. The BRS-transformation of $A^\mu_a$ and
$(\phi_a,\,H)$ has the form of an infinitesimal gauge transformation of the {\it un-shifted} fields. 
That is, $s\,A_a^\mu=D^\mu_{ab}\,u_b$ remains unchanged and
\beq
s\,\Phi=\frac{\kappa\,i}{2}\,\sigma_a\,u_a\>\Phi\ .
\eeq
The latter reads explicitly
\begin{gather}
s\,\phi_a=m\,u_a+\frac{\kappa}{2}\,(f_{abc}\,\phi_b\,u_c+H\,u_a)\ ,\notag\\
s\,H=-\frac{\kappa}{2}\,\phi_a\,u_a\ .
\end{gather}
It follows
\beq
s\,{\cal L}_{\rm YM}=0\ ,\quad s\,{\cal L}_{\Phi}=0\ .
\eeq
To keep $s^2\,A^\mu_a=0=s^2\, u_a$, the 
BRS-transformation of $u_a$ is unchanged: $ s\,u_a=-\frac{\kappa}{2}\,f_{abc}\,
u_b\,u_c$. With that one easily verifies $s^2\,\phi_a=0=s^2\, H$:
\beq
s^2\,\Phi=\frac{\kappa\,i}{2}\,\sigma_a\,(su_a)\>\Phi-
\frac{\kappa\,i}{2}\,\sigma_a\,u_a\>s\Phi=0\ ,
\eeq
due to $(\sigma_a\,u_a)\, (\sigma_b\,u_b)=i\,f_{abc}\,u_a\,u_b\,\sigma_c$.

The ghost Lagrangian is chosen such that $s\,({\cal L}_{\rm GF}+
{\cal L}_{\rm ghost})$ is a divergence: generalizing (\ref{L_ghost}) and 
(\ref{s(GF+ghost)}) one sets
\beq
{\cal L}_{\rm ghost}\=d \d_\mu\tilde u_a\>s\,A^\mu_a-m\,\tilde u_a\,s\,\phi_a
={\cal L}_{\rm ghost}^{(0)}+\kappa {\cal L}_{\rm ghost}^{(1)}\ ,
\eeq
which yields indeed
\beq
s\,({\cal L}_{\rm GF}+{\cal L}_{\rm ghost})=-\d_\mu\,I^\mu\ ,\quad
I^\mu\=d (\d_\nu A^\nu_a+m\phi_a)\,D^\mu_{ab}u_b=I^{(0)\,\mu}+\kappa\,\label{I:m>0}
I^{(1)\,\mu}\ .
\eeq
We end up with the total Lagrangian
\beq
{\cal L}_{\rm total}\=d {\cal L}_{\rm YM}+{\cal L}_\Phi+
{\cal L}_{\rm GF}+{\cal L}_{\rm ghost}=-\frac{\lambda\,v^4}{4}+{\cal L}^{(0)}+
\kappa\,{\cal L}^{(1)}+\kappa^2\,{\cal L}^{(2)}\ .
\eeq
The constant $-\frac{\lambda\,v^4}{4}$ is irrelevant and all terms of ${\cal L}^{(2)}$
come from ${\cal L}_{\rm YM}+{\cal L}_\Phi$. Note that ${\cal L}^{(0)}$ contains a 
divergence term, $-m\,\d_\mu(A^\mu_a\phi_a)$, which is irrelevant for the field equations 
but contributes to $j^{(n)\,\mu}$ (\ref{j^n}) and ${\cal L}^{(n)\,\mu}$ (\ref{L_1^n}).
For later purpose we give ${\cal L}^{(0)}$ explicitly:
\begin{gather}
{\cal L}^{(0)}=-\frac{1}{4}(A_a^{\mu,\nu}-A_a^{\nu,\mu})(A_{a\,\mu,\nu}-A_{a\,\nu,\mu})
-\frac{1}{2}(\d A_a)^2+\frac{m^2}{2}\,A^\mu_aA_{a\,\mu}\notag\\
+\d_\mu\tilde u_a\,\d^\mu u_a -m^2\,\tilde u_au_a\notag\\
+\frac{1}{2}\,\d_\mu\phi_a\,\d^\mu\phi_a-\frac{m^2}{2}\,\phi_a^2
+\frac{1}{2}\,\d_\mu H\,\d^\mu H-\lambda\, v^2\,H^2\notag\\
-m\,\d_\mu(A^\mu_a\,\phi_a)\ .\label{L_0:m>0}
\end{gather}
As in the massless case it is only the vanishing of $s^2\tilde u_a$ which relies on the 
field equations:
\beq
s^2\,A_a^\mu=0\ ,\quad s^2\,u_a=0\, ,\quad
s^2\tilde u_a=\frac{\delta\, S_{\rm total}}{\delta\,\tilde u_a}\ ,\quad
s^2\,\phi_a=0\ ,\quad s^2\, H=0
\eeq
and, hence, footnote \ref{fn:NL} applies also to the massive case.

Inserting (\ref{I:m>0}) and (\ref{L_0:m>0}) into our formulas 
(\ref{j^n}) (BRS-current) and (\ref{L_1^n}) ($Q$-vertex) we obtain
\begin{gather}
j^{(0)\,\mu}=(\d_\tau A^\tau_a+m\,\phi_a)\, \d^\mu u_a- (\d^\mu(\d_\tau A^\tau_a
+m\,\phi_a))\, u_a\notag\\
-\d_\nu\,\bigl((A^{\mu,\nu}_a-A^{\nu,\mu}_a)\,u_a\bigr)+((\square +m^2) A_a^\mu)\,u_a
\label{j:m>0}
\end{gather}
and
\begin{gather}
{\cal L}_1^{(1)\,\mu}=f_{abc}[A_{a\,\nu}\, u_b\,(\d^\mu A^\nu_c-\d^\nu A^\mu_c)
-\frac{1}{2}u_a\,u_b\,\d^\mu\tilde u_c+\frac{m}{2}\,A^\mu_a\,\phi_b\,u_c
-\frac{1}{2}\,\d^\mu\phi_a\,\phi_b\,u_c]\notag\\
+\frac{1}{2}\,\d^\mu H\,u_a\,\phi_a-\frac{1}{2}\,H\,u_a\,\d^\mu\phi_a
+\frac{m}{2}\,A^\mu_a\,u_a\,H\label{L_1:m>0}
\end{gather}
and again ${\cal L}_1^{(k)\,\nu}=0\ ,\>\forall k\geq 3$. For the same reason as in 
the massless case (\ref{K=0}) the terms in the second line of (\ref{j:m>0}) do not contribute
to $d_Q$. With that our results (\ref{j:m>0}) and (\ref{L_1:m>0}) agree with the literature 
\cite{DS,BD}.
\medskip

\noindent {\it Remarks: (1)} If one omits the divergence $-m\,\d_\mu(A^\mu_a\phi_a)$ in 
${\cal L}^{(0)}$ an additional term appears in $s\,{\cal L}_{\rm total}$: $I^\mu$
is replaced by $(I^\mu-m\,s\,(A^\mu_a\phi_a))$. With that our results for $j^{(0)}$
(\ref{j:m>0}) and ${\cal L}_1^{(1)}$ (\ref{L_1:m>0}) remain unchanged.

\noindent {(2)} For massive spin-1 fields $s_0\varphi_l$ is not a divergence 
(\ref{s_0=div}) for the basic fields $\varphi_l=\tilde u,\>\phi$. However, since the free field 
equations are differential equations of {\it second} order, we obtain $i\,(s_0\varphi_l)_{S_0}
=d_Q\,\varphi_{l\,S_0}\,,\>\varphi_l=\tilde u,\,\phi$, from part (ii) of Corollary 2.
\subsection{Massless spin-2 gauge fields}
In this Subsect.~we complete our proof of PGI-tree for massless spin-2 gauge fields: we
show that classical gravity can be formulated by a BRS-invariant Lagrangian 
${\cal L}_{\rm total}$ (\ref{sL1}) which fits in our formalism. To satisfy the latter,
${\cal L}_{\rm total}$ must be a (formal) power series in $\kappa$ and it must be a 
polynomial only in zeroth and first derivatives of the basic fields. (It will turn out that both 
properties are non-trivial.) Such a BRS-formulation of gravity was given
by Kugo and Ojima in Sect.~2 of \cite{KO}. 
In that formalism we choose Feynman gauge $\alpha_0=1$
and eliminate the Nakanishi-Lautrup field $b_\mu$ \cite{LN} 
(by inserting the field equation for $b_\mu$). In view of
perturbation theory around the Minkowski metric $\eta^{\mu\nu}$
we introduce a field $h^{\mu\nu}$ which is the deviation from $\eta^{\mu\nu}$,
in terms Goldberg variables $\tilde g^{\mu\nu}$ (for details see e.g. \cite{S-wiley}):
\beq
\tilde g^{\mu\nu}(x)\=d \sqrt{-g(x)}\,g^{\mu\nu}(x)=\eta^{\mu\nu}+\kappa\,h^{\mu\nu}(x)\ .\label{h}
\eeq
The inverse tensor is a kind of geometric series in $\kappa h$ (which we understand 
as formal power series in $\kappa$):
\beq
\tilde g_{\mu\nu}\=d \frac{1}{\sqrt{-g}}\,g_{\mu\nu}=\eta_{\mu\nu}-\kappa\,h_{\mu\nu}
+\kappa^2\,h_{\mu\alpha}h^\alpha_{\>\nu}-...\ ,\label{h1}
\eeq
where
\beq
h^\alpha_{\>\nu}=\eta_{\nu\tau}\, h^{\alpha\tau}\ ,\quad
h_{\mu\nu}=\eta_{\mu\rho}\,\eta_{\nu\tau}\,h^{\rho\tau}\ ,
\eeq
and we set $h\=d h^\mu_\mu$. The field algebra ${\cal P}$ is the polynomial algebra 
generated by $h^{\mu\nu}$, the fermionic vector ghost fields $u^\mu,\>\tilde u_\mu$
and all partial derivatives of these fields. (The indices of $u$ and $\tilde u$ are also raised 
and lowered by means of $\eta$.) 
According to \cite{KO} the BRS-transformation is of the form $s=s_0+\kappa\, s_1$
and on  the basic fields it is given by
\begin{gather}
s\,h^{\mu\nu}=u^{\nu,\mu}+u^{\mu,\nu}-\eta^{\mu\nu}\,u^\rho_{,\rho}
+\kappa\,(h^{\mu\sigma}\,u^\nu_{\>,\sigma}+h^{\nu\sigma}\,u^\mu_{\>,\sigma}-
(h^{\mu\nu}\,u^\rho)_{,\rho})\ ,\notag\\
s\,u^\mu=-\kappa\,(u^\lambda\,u^\mu_{\>,\lambda})\ ,\notag\\
s\,\tilde u_\mu=-h_{\mu\rho}^{\>\>,\rho}\ .\label{s:gravity}
\end{gather}
(The sign of $s\,u^\mu$ is determined by the requirement $s^2\,h^{\mu\nu}=0$.) By using
\beq
D^{\mu\nu}_{\>\rho}\=d (\eta^{\mu\sigma}+\kappa\,h^{\mu\sigma})\,\delta^\nu_\rho\,\d_\sigma
+(\eta^{\nu\sigma}+\kappa\,h^{\nu\sigma})\,\delta^\mu_\rho\,\d_\sigma
-\d_\rho\bigl((\eta^{\mu\nu}+\kappa\,h^{\mu\nu})\>\>\cdot\>\>\bigr)
\eeq
we may also write
\beq
s\,h^{\mu\nu}=D^{\mu\nu}_{\>\rho}\,u^\rho\ .
\eeq

In terms of the Christoffel symbols 
\beq
\Gamma^\alpha_{\beta\gamma}=\frac{1}{2}\,g^{\alpha\mu}\,(g_{\beta\mu,\gamma}
+g_{\mu\gamma,\beta}-g_{\beta\gamma,\mu})\label{Gamma}
\eeq
the Einstein-Hilbert Lagrangian reads
\beq
{\cal L}_{\rm E}=\frac{1}{\kappa^2}\,\sqrt{-g}\,g^{\mu\nu}\,(\Gamma^\rho_{\nu\rho,\mu}-
\Gamma^\rho_{\mu\nu,\rho}+\Gamma^\rho_{\mu\sigma}
\,\Gamma^\sigma_{\nu\rho}-\Gamma^\sigma_{\mu\nu}\,
\Gamma^\rho_{\sigma\rho})\ .
\eeq
But ${\cal L}_{\rm E}$ is unsuitable for our formalism: it contains terms $\sim\kappa^{-1}$
and second derivatives of $h^{\mu\nu}$. Both shortcomings can be
removed by subtracting a divergence \cite{KO}:
\begin{gather}
{\cal L}'_{\rm E}\=d {\cal L}_{\rm E}-\d_\mu {\cal D}^\mu\ ,\\
{\cal D}^\mu\=d \frac{1}{\kappa^2}\,(\tilde g^{\mu\nu}\,\Gamma^\lambda_{\nu\lambda}-
\tilde g^{\rho\sigma}\,\Gamma^\mu_{\rho\sigma})=\frac{1}{\kappa^2}\,
\bigl(\frac{1}{2}\,\tilde g^{\mu\nu}\,\tilde g_{\alpha\beta}\,\d_\nu\tilde g^{\alpha\beta}\,
+\d_\nu\tilde g^{\mu\nu}\bigr)\notag\\
=\frac{1}{\kappa}\,{\cal D}^{(-1)\,\mu}+{\cal O}(\kappa^0)\ ,\quad {\cal D}^{(-1)\,\mu}=
\frac{1}{2}\,h^{,\mu}+h^{\mu\nu}_{\> ,\nu}\ .
\end{gather}
By inserting (\ref{h})-(\ref{h1}) and (\ref{Gamma}) we indeed obtain a formal power series 
in $\kappa$ for ${\cal L}'_{\rm E}$:
\beq
{\cal L}'_{\rm E}=\sum_{n=0}^\infty\kappa^n\,{\cal L}^{\prime\,(n)}_{\rm E}\ ,\label{L'_E}
\eeq
see Sect.~5.5 of \cite{S-wiley}. 
The total Lagragian ${\cal L}_{\rm total}$ of the BRS-formalism
is obtained by adding a gauge fixing term ${\cal L}_{\rm GF}$ and a ghost
term ${\cal L}_{\rm ghost}$:
\begin{gather}
{\cal L}_{\rm total}={\cal L}'_{\rm E}+{\cal L}_{\rm GF}+{\cal L}_{\rm ghost}=
\sum_{n=0}^\infty\kappa^n\,{\cal L}^{(n)}
\ ,\label{L_total:gravity}\\
{\cal L}_{\rm GF}=\frac{1}{2}\,h^{\alpha\beta}_{\>,\beta}\,h_{\alpha\rho}^{\>,\rho}=
{\cal L}_{\rm GF}^{(0)}\ ,\\
{\cal L}_{\rm ghost}=-\frac{1}{2}\,(\tilde u_{\nu,\mu}+\tilde u_{\mu,\nu})\>
s\,h^{\mu\nu}={\cal L}_{\rm ghost}^{(0)}+\kappa\,{\cal L}^{(1)}_{\rm ghost}\ .
\end{gather}
With that the BRS-transformation is nilpotent modulo the field equations:
\beq
s^2\,h^{\mu\nu}=0\ ,\quad s^2\,u^\mu=0\ ,\quad 
s^2\,\tilde u_\mu=-\eta_{\mu\tau}\,(D^{\tau\nu}_\rho\,u^\rho)_{,\nu}=
-\frac{\delta\,S_{\rm total}}{\delta\,\tilde u^\mu}\ .\label{s^2(tilde-u)}
\eeq

We turn to the verification of the BRS-invariance of ${\cal L}_{\rm total}$.
Again, the BRS-transformation of the gauge field
$h^{\mu\nu}$ has the form of an infinitesimal gauge transformation 
(i.e.~general coordinate transformation) and, hence,
$s\,{\cal L}_{\rm E}$ is known from the gauge variation of ${\cal L}_{\rm E}$ \cite{KO}:
\beq
s\,{\cal L}_{\rm E}=-\d_\mu(\kappa\,{\cal L}_{\rm E}\,u^\mu)\ .\label{sL_E}
\eeq
By the same calculation as in (\ref{s(GF+ghost)}) we obtain
\beq
s\,({\cal L}_{\rm GF}+{\cal L}_{\rm ghost})=\d_\mu F^{\mu}\ ,\quad 
F^\mu\=d h_{\nu\rho}^{\>,\rho}\,D^{\mu\nu}_{\>\lambda}\,u^\lambda
=F^{(0)\,\mu}+\kappa\, F^{(1)\,\mu}\ .\label{s(L_GF+L_FP)}
\eeq
Summing up we get
\beq
s\,{\cal L}_{\rm total}=-\d_\mu\,I^\mu\ ,\quad 
I^\mu\=d \kappa\,{\cal L}_{\rm E}\,u^\mu+s\,{\cal D}^\mu-F^\mu\ 
+\frac{1}{\kappa}\,\d_\rho (u^{\rho,\mu}-u^{\mu,\rho})=
\sum_{n=0}^\infty \kappa^n\, I^{(n)\,\mu}\ .\label{sL:gravity}
\eeq
In $I^\mu$ we have added the conserved vector field
$\frac{1}{\kappa}\,\d_\rho (u^{\rho,\mu}-u^{\mu,\rho})$
to cancel $s_0\,{\cal D}^{(-1)\,\mu}=\square u^\mu-u^{\rho\>,\mu}_{\>,\rho}$,
because in our formalism $I^\mu$ is assumed to be a formal power 
series in $\kappa$.

By means of our formula (\ref{j^n}) we compute the zeroth order of the BRS-current.
We obtain
\beq
j_{S_0}^{(0)\,\mu}=-h^{\alpha\beta}_{S_0\>,\beta}\,\d^\mu u_{S_0\>,\alpha}+
(\d^\mu h^{\alpha\beta}_{S_0\>,\beta})\,u_{S_0\>,\alpha}
+\d_\rho(K_{S_0}^{\mu\rho}-K_{S_0}^{\rho\mu})\ ,\label{j^0-p}
\eeq
where 
\beq
K^{\rho\mu}\=d \frac{1}{2}\,h^{,\rho}\,u^\mu +h^{\alpha\rho,\mu}\,u_\alpha +
h^{\rho\nu}_{\>,\nu}\,u^\mu+h^{\rho\nu}\,u^\mu_{\>,\nu}\ ,
\eeq
for details see Appendix B. As explained in (\ref{K=0}), the term
$\d_\rho(K^{\mu\rho}-K^{\rho\mu})$ does not contribute to $d_Q$. The other terms 
are precisely the terms which are used for the construction of $d_Q$ in 
\cite{Schorn,S-wiley,G:grav}.
\section{Outlook}\setcounter{equation}{0}
What do we learn from this paper with respect to model building, i.e.~the task:
\begin{itemize}
\item given a free BRS invariant theory $({\cal L}^{(0)},s_0)$ which satisfies (\ref{Q-vaccum}),
find a non-trivial deformation ${\cal L}^{(0)}\rightarrow {\cal L}_{\rm total}=
{\cal L}^{(0)}+{\cal L}_{\rm int}$ (\ref{L_total}) such that ${\cal L}_{\rm total}$
satisfies PGI-tree (and some obvious further conditions, e.g.~Lorentz invariance,
${\cal L}_{\rm total}^*={\cal L}_{\rm total}$ and ghost number zero)?
\end{itemize}
In the literature this problem is usually treated by making a polynomial ansatz for 
${\cal L}_{\rm int}$ and working out the condition of PGI-tree, as mentioned in point (B) 
of the introduction. Due to this paper one can proceed alternatively, namely one searches 
for deformations ${\cal L}^{(0)}\rightarrow {\cal L}_{\rm total}$ and $s_0\rightarrow s$
(with (\ref{s}), (\ref{sdphi}) and (\ref{s^2=0})) such that $s{\cal L}_{\rm total}$ is a divergence
(\ref{sL1}). This amounts to an inductive determination of the sequences 
$({\cal L}^{(n)})_n$ and $(s_n)_n$. (In Sect.~5.3 and Appendix B of \cite{DF2}
and in \cite{Hurth} analogous procedures are given to solve the local current conservation 
(\ref{dj(g)}).)  It is not yet investigated, whether this procedure yields the {\it most general}
solution of the above task, but it would be rather surprising if there
were additional solutions. 
\medskip

The method of proof given in this paper applies not only to BRS-symmetry; all 
Lagrangians ${\cal L}_{\rm total}$ and infinitesimal symmetry transformations $s$ which 
satisfy (\ref{L_total})-(\ref{sdphi}) and (\ref{Q-vaccum}) (where $d_Q$ is constructed by
(\ref{d_Q1}) from the zeroth order $j^{(0)\,\mu}$ (\ref{j^n})
of the Noether current belonging to $({\cal L}_{\rm total},s)$) are admitted. 
Since the nilpotency (\ref{s^2=0})
of the BRS-transformation is not used in our proof, the restrictions on $s$
are not strong. A trivial example are global $U(1)$-symmetries, e.g. charge 
conservation in QED: for the spinors $\psi$ and $\bar\psi$ let $s\psi=i\psi$,
$s\bar\psi=-i\bar\psi$, and the photon field $A^\mu$ is uncharged, $sA^\mu=0$.
The $s$-variation of the Lagrangian of QED, ${\cal L}^{\rm QED}_{\rm total}=
{\cal L}^{(0)}+\kappa g(x){\cal L}^{(1)}$, even vanishes:
$s {\cal L}^{\rm QED}_{\rm total}=0$. Our method yields
\beq
\Bigl[Q_\psi , T^{\rm tree}_{n\>S_0}\Bigl((\int g\,{\cal L}^{(1)})^{\otimes n}\Bigr)\Bigr]=0\ ,
\eeq
where $Q_\psi\=d\int d^3x\,(\bar\psi\gamma^0\psi)_{S_0}$, cf. Appendix B of \cite{DF}.
\medskip

Can the method of this paper be applied to {\it massive} spin-2 gauge fields?
This amounts to the question whether there is a BRS-invariant Lagrangian
for such fields with the properties (\ref{L_total})-(\ref{sdphi}) and (\ref{Q-vaccum}).

In \cite{GS} the requirement of PGI-tree has been applied to {\it massive} spin-2 fields.
Starting with the most reasonable free theory (in Feynman gauge) and proceeding 
similarly to \cite{SW,S-wiley} the lowest orders 
${\cal L}^{(1)}$ and ${\cal L}^{(2)}$ of the possible interaction ${\cal L}=\sum_{n=1}^\infty
\kappa^n\,{\cal L}^{(n)}$ have been derived. The result agrees with 
the corresponding terms of the Einstein-Hilbert Lagrangian with cosmological 
constant $\Lambda\not= 0$:
${\cal L}_{E\,\Lambda}:={\cal L}_E-\frac{2}{\kappa^2}\,\sqrt{-g}\,\Lambda$. 
The mass $m$ 
of the free spin-2 field is related to $\Lambda$ by $m^2=2\,\Lambda$. 

${\cal L}_{E\,\Lambda}$ satisfies the BRS-invariance (\ref{sL_E}), too.
But the expansion of  ${\cal L}_{E\,\Lambda}$ in powers of $\kappa$ 
contains a term $\frac{-1}{\kappa}
\,\Lambda\,h$, which cannot be removed by adding a divergence, and the field equation to 
order $\kappa^{-1}$ is nonsense: $\Lambda=0$. (This contradiction is due to the fact that the 
Minkowski metric $\eta^{\mu\nu}$ is not a solution of the field equation to
${\cal L}_{E\,\Lambda}$.) Therefore, it seems that
the formalism of this paper cannot be applied to ${\cal L}_{E\,\Lambda}$;
omission of  the term $\frac{-1}{\kappa}\,\Lambda\,h$ destroys the BRS-invariance
of ${\cal L}_{E\,\Lambda}$.

It is doubtful whether the mass can be generated  
via Higgs mechanism (as for spin-1 fields), since we are not aware of such a 
procedure in the literature and since in \cite{GS} it has turned out that PGI-tree can be 
satisfied up to second order without introducing a physical Higgs field.
\medskip

\begin{center}
{\Large\bf Appendices}
\end{center}

\begin{appendix}
\section{Algebraic off-shell formalism}\setcounter{equation}{0}
This Appendix is a short introduction to the formalism given in \cite{DF2} and \cite{DF3}.
Let $\varphi$ be a complex scalar field in 4 dimensions.
The classical phase space is ${\cal C}\=d \mathcal{C}^{\infty}(\RR^4,\CC)$.
The classical field $(\d^a\varphi)(x)$, $a\in\NN_0^4$, is the evaluation functional
\beq
(\d^a\varphi) (x)\>:\>{\cal C}\longrightarrow\CC\>,\>
(\d^a\varphi) (x)(h)=\d^ah(x)\ ,\quad \varphi^\star (x)(h)=\overline{h(x)}\ .\label{eval-func}
\eeq
Let $\mathcal{F}$ be the set of all functionals\footnote{$\CC[[\hbar]]$ is the space of all
formal power series $\sum_{l=0}^\infty c_l\hbar^l$ with $c_l\in\CC$.}
$F\equiv F(\varphi):{\cal C}\longrightarrow\CC[[\hbar]]$ which have the form
\begin{equation}
  F(\varphi)=\sum_{n=0}^N\int dx_1\ldots dx_n\,\varphi(x_1)
  \cdots\varphi(x_n)f_n(x_1,\ldots,x_n)\ ,\quad\quad N<\infty\ ,
  \label{F(phi)} 
\end{equation}
where $f_0\in\CC[[\hbar]]$. The higher $f_n$'s are $\CC[[\hbar]]$-valued 
distributions with compact support, which are symmetric under 
permutations of $x_1,...,x_n$ and which are 'translation invariant up to smooth
functions', i.e.
\begin{equation}
        WF(f_{n})\subset \{(x,k)\>|\>\sum_{i=1}^n k_{i}=0\}\ .
\end{equation}
Due to (\ref{eval-func}) it holds
$F(\varphi)(h)=F(h)$. $\mathcal{F}$ is a *-algebra with
the classical product $(F_1\cdot F_2)(h):=F_1(h)\cdot F_2(h)$
and $ \langle f_{n},\varphi^{\otimes  n}\rangle ^\star=  \langle \overline{f_n},
(\varphi^\star)^{\otimes n}\rangle$. The functional
$\omega_0:\mathcal{F}\rightarrow \CC[[\hbar]],\>\omega_0(\sum_n
\langle f_{n},\varphi^{\otimes  n}\rangle)=f_0$ is the 'vacuum state'. 

The support of $F\in \mathcal{F}$ is the support of $\frac{\delta F}{\delta \varphi}$.
Let ${\cal P}$ be the algebra of all polynomials in $\varphi,\varphi^*$ and their partial 
derivatives. The vector space $\mathcal{F}_{\text{loc}}$ of {\it local} functionals
is the set of all $F\in {\cal F}$ of the form
\begin{equation}
  F=\int dx\,\sum_{i=1}^N A_i(x)h_i(x)\ ,\quad A_i\in {\cal P}\ ,\> h_i\in {\cal D}(\RR^4)\ .
\end{equation}

Given an action $S$, the set ${\cal C}_{S}\subset {\cal C}$ is the set of all smooth 
solutions of the Euler-Lagrange equations belonging to $S$
with compactly supported Cauchy data. We set
$F_{S}\=d F\vert_{{\cal C}_{S}}\ ,\>F\in {\cal F}$. We study actions of the form
$S_{\rm total}=S_0+S_{\rm int}$ with free part $S_0$ and compactly supported
interacting part: $S_{\rm int}\in {\cal F}_{\text{loc}}$. We always assume that 
$S_{\rm total}$ is such that the Cauchy problem has a unique solution.
Given two actions $S_{\rm total}^{(j)}=S_0+S_{\rm int}^{(j)}\> (j=1,2)$ of this kind, there 
exists a map
\begin{equation}
   r_{S_{\rm total}^{(1)},S_{\rm total}^{(2)}}\>:\>
\mathcal{C}_{S_{\rm total}^{(2)}}\rightarrow
   \mathcal{C}_{S_{\rm total}^{(1)}}\ ,\ f_2\mapsto f_1\ ,
\end{equation}
such that $f_1$ agrees with $f_2$ outside the future of $(\supp 
\frac{\delta S_{\rm int}^{(1)}}{\delta \varphi}\cup\supp
\frac{\delta S_{\rm int}^{(2)}}{\delta \varphi})$. A {\it retarded field}
\beq
A^{\rm ret}_{S_{\rm int}}(x)\=d A(x)\circ r_{S_0+S_{\rm int},S_0}
\>:\>{\cal C}_{S_0}\longrightarrow\CC\ ,\quad A\in {\cal P}\ ,\label{A^ret:def}
\eeq
is a functional on the free solutions which solves the field equation
for $S_{\rm total}=S_0+S_{\rm int}$ if $A=\varphi$ or $\varphi^*$. The 
perturbation expansion of a classical interacting field is the Taylor 
series of the corresponding retarded field as functional of $S_{\rm int}$,
\beq
A^{\rm ret}_{S_{\rm int}}(x)\simeq \sum_{n=0}^\infty\frac{1}{n!}
R^{\rm class}_{n,1\>S_0}(S_{\rm int}^{\otimes n},A(x))\equiv :
R^{\rm class}_{S_0}\Bigl(e_\otimes^{S_{\rm int}},A(x)\Bigr)\ ,\label{A^ret}
\eeq
where $R^{\rm class}_{n,1\>S_0}: {\cal F}_{\text{loc}}^{\otimes (n+1)}
\rightarrow {\cal F}\vert_{{\cal C}_{S_0}}$
is the classical retarded product. (The lower index $S_0$ of $R^{\rm class}_{S_0}$
is redundant, however we write it in view of the conventions in QFT.)

Next we study the quantization of the free theory. 
Let $\Delta^{(m)}_+$ be the 2-point function of the free scalar field
with mass $m$. We define a $*$-product $\star\equiv\star_m$ on $\mathcal{F}$
\cite{BFFLS},
\begin{gather}
  (F\star_m G)(\varphi)=
  \sum_{n=0}^{\infty}\frac{\hbar^n}{n!}
  \int dx_1\ldots dx_n dy_1\ldots dy_n 
  \frac{\delta^n F}{\delta\varphi(x_1)\cdots\delta\varphi(x_n)}\notag\\
  \cdot \prod_{i=1}^n \Delta^{(m)}_+(x_i-y_i) 
  \frac{\delta^n G}{\delta\varphi(y_1)\cdots\delta\varphi(y_n)}\ ,
  \label{*-product}
\end{gather}
which is associative and non-commutative.
Let $\mathcal{J}^{(m)}\subset {\cal F}$ be the ideal (with respect to the classical product)
generated by the free field equation $(\square +m^2)\varphi =0$.
Let $\mathcal{F}^{(m)}_{0}\equiv {\cal F}/{\cal J}^{(m)}$. Due to $(\square +m^2)
\Delta^{(m)}_+=0$ the $*$-product (\ref{*-product}) induces a well defined product 
$\mathcal{F}^{(m)}_{0}\times \mathcal{F}^{(m)}_{0}\rightarrow\mathcal{F}^{(m)}_{0}$,
and we denote the corresponding algebra by ${\cal A}_0^{(m)}\equiv
(\mathcal{F}^{(m)}_{0},\star_m)$. The latter can
be faithfully represented on Fock space, the $*$-product goes over into
the operator product and, in addition, the classical product into the normally 
ordered product. $\omega_0$ induces a state on
$\mathcal{A}^{(m)}_{0}$ which corresponds to the Fock vacuum.

Finally we turn to perturbative QFT.
Let $F,S_n\in {\cal F}_{\rm loc}$ with $F,S_n\sim\hbar^0$ $(n\in {\bf N})$, and
let $S\equiv S(\kappa)=\sum_{n=1}^{\infty}\kappa^n S_n$ be a formal power series 
with $S(0)=0$.
We associate to $(F,S)$ a formal power series 
\beq
F_{S/\hbar}=\sum_{n=0}^\infty\frac{1}{n!}
R_{n,1}((S/\hbar)^{\otimes n},F)\equiv :R(e_\otimes^{S/\hbar},F)\label{intfield:QFT}
\eeq
which we interpret as the functional $F$ of the 
interacting retarded field under the influence of 
the interaction $S$ where $\kappa$ is the expansion parameter 
of the formal power series. The retarded products
\beq
R_{n,1}\>:\>{\cal F}_{\rm loc}^{\otimes (n+1)}\longrightarrow {\cal F}\label{ret-prod:QFT}
\eeq
are {\it linear} maps which are {\it symmetric in the first $n$ factors}.
Their basic properties are
\begin{itemize}
 \item zeroth order $\quad R_{0,1}(F)=F$\ ,
\item Causality
\begin{equation}
\supp R_{n,1}\subset
\{(x_1,\ldots,x_{n},x)\in\RR^{4(n+1)}\> |\> x_i\in x+\overline{V}_-,\>
\forall i=1,\ldots n\}\label{caus}
\end{equation}
\item and the GLZ relation
\begin{equation}
i\Bigl[R\Bigl(e_\otimes^{S/\hbar},F\Bigr),R\Bigl(e_\otimes^{S/\hbar},H\Bigr)\Bigr]_\star =
R\Bigl(e_\otimes^{S/\hbar}\otimes H,F\Bigr)-(H\leftrightarrow F)\ .\label{GLZ}
\end{equation}
\end{itemize}
We set $R_{S_0}(...)\=d R(...)\vert_{{\cal C}_{S_0}}$. 

$R_{n,1}^{\rm tree}$ is the contribution of the tree diagrams to $R_{n,1}$. 
As shown in Sect.~5.2 of \cite{DF1}, $R_{n,1}^{\rm tree}$ is that
part of $R_{n,1}$ with the lowest power of $\hbar$, more precisely
\beq
R_{n,1}(F_1,...,F_{n+1})=R^{\rm tree}_{n,1}(F_1,...,F_{n+1})+
{\cal O}(\hbar^{n+1})\quad\mathrm{and}\quad
R^{\rm tree}_{n,1}(F_1,...,F_{n+1})\sim\hbar^{n}\label{R-tree}
\eeq
if $F_1,...,F_{n+1}\sim\hbar^0$.
\section{Zeroth order of the BRS-current for spin-2 gauge fields}\setcounter{equation}{0}
From ${\cal L}^{(0)}$ (\ref{L_total:gravity}) one obtains the free field equations
\beq
\square\, h^{\mu\nu}_{S_0}=0\ ,\quad \square\, u^{\mu}_{S_0}=0\ ,\quad 
\square\, \tilde u_{\mu\,S_0}=0\ .\label{ffeq}
\eeq
By using (\ref{j^n}) and (\ref{sL:gravity}) we obtain
\beq
 j^{(0)\,\mu}=-\frac{\d{\cal L}^{(0)}}{\d\varphi_{i,\mu}}\,s_0\,\varphi_i-
\d_\nu\,{\cal D}^{(-1)\,\nu}\,u^\mu-s_0\,{\cal D}^{(0)\,\mu}-s_1\,{\cal D}^{(-1)\,\mu}
+F^{(0)\,\mu}\ ,\label{j^0}
\eeq
where we have 
taken into account ${\cal L}_{\rm E}^{(-1)}=\d_\mu\,{\cal D}^{(-1)\,\mu}$.
After restriction to ${\cal C}_{S_0}$ (\ref{ffeq}), we obtain the following non-vanishing 
contributions to $(-j^{(0)\,\mu}_{S_0})$:
\begin{gather}
\Bigl(\frac{\d{\cal L}^{\prime\,(0)}_{\rm E}}{\d h^{\alpha\beta}_{\>,\mu}}
\>s_0\,h^{\alpha\beta}\Bigr)_{S_0}=\frac{1}{2}\,h^{\>,\mu}_{\alpha\beta\,S_0}\,
(u^{\alpha,\beta}_{S_0}+u^{\beta,\alpha}_{S_0})-
h^{\alpha\mu,\beta}_{S_0}\,(u_{\alpha,\beta\,S_0}+u_{\beta,\alpha\, S_0}
-\eta_{\alpha\beta}\,u^{\lambda}_{\>,\lambda\,S_0})\ ,\notag\\
\Bigl(\frac{\d{\cal L}_{\rm GF}}{\d h^{\alpha\beta}_{\>,\mu}}
\>s_0\,h^{\alpha\beta}\Bigr)_{S_0}=h^{\>,\gamma}_{\alpha\gamma\,S_0}\,
(u^{\alpha,\mu}_{S_0}+u^{\mu,\alpha}_{S_0}-
\eta^{\alpha\mu}\,u^{\lambda}_{\>,\lambda\,S_0})\ ,\notag\\
\Bigl(\frac{\d{\cal L}^{(0)}_{\rm ghost}}{\d \tilde u_{\nu,\mu}}
\>s_0\,\tilde u_{\nu}\Bigr)_{S_0}=(u^{\mu}_{\>,\tau\,S_0}+u^{\>,\mu}_{\tau\,S_0})\,
h^{\tau\rho}_{\>,\rho\,S_0}-u^{\lambda}_{\>,\lambda\,S_0}\,
h^{\mu\rho}_{\>,\rho\,S_0}\ ,\notag\\
\bigl(\d_\nu\,{\cal D}^{(-1)\,\nu}\bigr)_{S_0}\,u_{S_0}^{\mu}=
h^{\alpha\beta}_{\>,\alpha\beta\,S_0}\,u^{\mu}_{S_0}\,\notag\\
\bigl(s_0\,{\cal D}^{(0)\,\mu}+s_1\,{\cal D}^{(-1)\,\mu}\bigr)_{S_0}=
\frac{1}{2}\,h_{,\nu\,S_0}\,(u^{\mu,\nu}_{S_0}+u^{\nu,\mu}_{S_0})+\frac{1}{2}\,
h_{S_0}\,u^{\lambda,\mu}_{\>,\lambda\,S_0}-\frac{1}{2}\,(h_{S_0}\,
u^{\lambda}_{S_0})^{,\mu}_{,\lambda}\notag\\
-h^{\mu\nu}_{S_0}\,u^{\lambda}_{\>,\lambda\nu\,S_0}
+(h^{\rho\nu}_{S_0}\,u^{\mu}_{\>,\nu\,S_0})_{,\rho}
-(h^{\mu\nu}_{\>,\nu\,S_0}\,u^{\rho}_{S_0})_{,\rho}\,\notag\\
-\bigl(F^{(0)\,\mu}\bigr)_{S_0}=-h^{\tau\nu}_{\>,\nu\,S_0}\,u^{\mu}_{\>,\tau\,S_0}
-h^{\tau\nu}_{\>,\nu\,S_0}\,u^{\>,\mu}_{\tau\,S_0}+
h^{\mu\nu}_{\>,\nu\,S_0}\,u^{\lambda}_{\>,\lambda\,S_0}\ .
\end{gather}
The sum of these terms is equal to (\ref{j^0-p}).

\end{appendix}

\vskip0.5cm
{\bf Acknowledgments:} Discussions with Klaus Fredenhagen and G\"unter Scharf have  
been very helpful and motivating for writing this paper.

\end{document}